\documentclass[aps,prb,amsmath, superscriptaddress, amssymb,preprint]{revtex4-1}

\usepackage{color}
\usepackage{rotating}
\usepackage{graphicx}
\usepackage{dcolumn}
\usepackage{bm}
\usepackage{float}
\usepackage{verbatim}
\usepackage{makecell}
\usepackage{array}
\usepackage{setspace}
\usepackage{mathptmx}
\usepackage{CJKutf8}
\usepackage{xr}

\usepackage{xr}
\makeatletter
\newcommand*{\addFileDependency}[1]{
  \typeout{(#1)}
  \@addtofilelist{#1}
  \IfFileExists{#1}{}{\typeout{No file #1.}}
}
\makeatother

\newcommand*{\myexternaldocument}[1]{
    \externaldocument{#1}
    \addFileDependency{#1.tex}
    \addFileDependency{#1.aux}
}

\myexternaldocument{main_SI}

\let\baraccent=\= 
\renewcommand{\=}[1]{\stackrel{#1}{=}} 

\newcommand{\ab}[2]{\mspace{-4.0mu}\left(\mspace{-8.0mu}
\begin{smallmatrix}&\ifthenelse{\equal{#1}{}}{a}{#1} \\&\ifthenelse
{\equal{#2}{}}{b}{#2}\end{smallmatrix}\mspace{-3.0mu}\right)}

\newcommand{\kv}{\mspace{-4.0mu}\left(\mspace{-8.0mu}
\begin{smallmatrix}&\mathbf{\pmb{\kappa}} \\&\nu\end{smallmatrix}
\mspace{-3.0mu}\right)}

\newcommand{\angstrom}{\mbox{\normalfont\AA}}

\begin{document}
\preprint{}

\title{Phonon Confinement and Transport in Ultrathin Films}

\author{Bo Fu\begin{CJK*}{UTF8}{gbsn}
(傅博)
\end{CJK*}}
\affiliation{MOE Key Laboratory of Thermo-Fluid Science and Engineering, School of Energy and Power Engineering, Xi'an Jiaotong University, Xi'an 710049, China}
\affiliation{Department of Mechanical Engineering, Carnegie Mellon University, Pittsburgh, PA 15213, USA}
\author{Kevin D. Parrish}
\affiliation{Department of Mechanical Engineering, Carnegie Mellon University, Pittsburgh, PA 15213, USA}
\author{Hyun-Young Kim}
\affiliation{Department of Mechanical Engineering, Carnegie Mellon University, Pittsburgh, PA 15213, USA}
\author{Guihua Tang\begin{CJK*}{UTF8}{gbsn} (唐桂华) \end{CJK*}} \email{ghtang@mail.xjtu.edu.cn.}
\affiliation{MOE Key Laboratory of Thermo-Fluid Science and Engineering, School of Energy and Power Engineering, Xi'an Jiaotong University, Xi'an 710049, China}
\author{Alan J. H. McGaughey} \email{mcgaughey@cmu.edu.}
\affiliation{Department of Mechanical Engineering, Carnegie Mellon University, Pittsburgh, PA 15213, USA}

\date{\today}

\begin{abstract}
Thermal transport by phonons in films with thicknesses of less than 10 nm is investigated in a soft system (Lennard-Jones argon) and a stiff system (Tersoff silicon) using two-dimensional lattice dynamics calculations and the Boltzmann transport equation.
This approach uses a unit cell that spans the film thickness, which removes approximations related to the finite cross-plane dimension required in typical three-dimensional-based approaches.
Molecular dynamics simulations, which make no assumptions about the nature of the thermal transport, are performed to obtain finite-temperature structures for the lattice dynamics calculations and to predict thermal conductivity benchmarks.
Thermal conductivity decreases with decreasing film thickness for both the two-dimensional lattice dynamics calculations and the MD simulations, until the thickness reaches four unit cells (2.1 nm) for argon and three unit cells (1.6 nm) for silicon.
With a further decrease in film thickness, thermal conductivity plateaus in argon while it increases in silicon. This unexpected behavior, which we identify as a signature of phonon confinement, is a result of an increased contribution from low-frequency phonons, whose density of states increases as the film thickness decreases.
Phonon mode-level analysis suggests that confinement effects emerge below thicknesses of ten unit cells (5.3 nm) for argon and six unit cells (3.2 nm) for silicon. These transition points both correspond to approximately twenty atomic layers.
Thermal conductivity predictions based on the bulk (i.e., three-dimensional) phonon properties combined with a boundary scattering model do not capture the low thickness behavior.
To match the two-dimensional lattice dynamics and molecular dynamics predictions for larger thicknesses, the three-dimensional lattice dynamics calculations require a finite specularity parameter that in some cases approaches unity.
These findings point to the challenges associated with interpreting experimental thermal conductivity measurements of ultrathin silicon films, where surface roughness and a native oxide layer impact phonon transport.
\end{abstract}

\maketitle

\section{\label{sec:intro}Introduction} 

Advances in nano-fabrication have pushed the minimum feature size in electronic, optoelectronic, and energy conversion devices to below 10 nm.\cite{heremans2013thermoelectrics, ieong2004silicon, wang2012electronics} Thin films are prototypical nanostructures that are used in thermoelectric energy conversion, \cite{venkatasubramanian2001thin} solar energy conversion, \cite{wang2014device} semiconductor laser,\cite{lu2012plasmonic} and solid state lighting\cite{ying2014white} devices. The thermal transport properties of a film can vary dramatically compared to its corresponding bulk material.\cite{cahill2003nanoscale, cahill2014nanoscale} In an electrically-insulating or semiconducting material, the dominate heat carriers are phonons, which scatter more often with a film's boundaries as its thickness is reduced. The resulting thermal conductivity reduction is problematic when heat dissipation is important (e.g., in a laser), but desirable in thermoelectric energy conversion.

\begin{figure}[tb]
\centering
\includegraphics{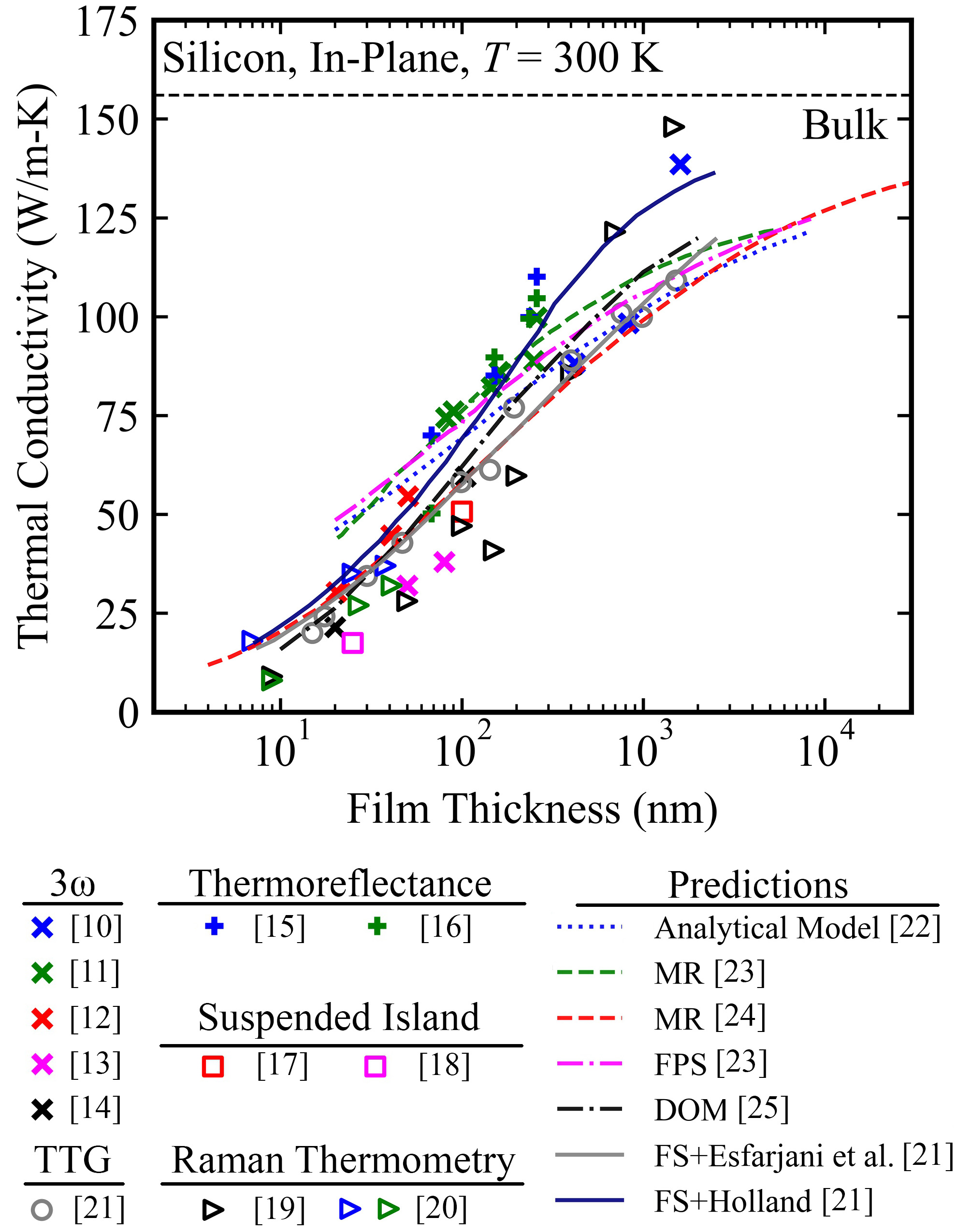} 
\caption{Silicon film in-plane thermal conductivity data from experimental measurements (markers) and predictions (lines) at a temperature of 300 K. The embedded films (cross/plus markers) were measured by $\rm 3-\omega$ \cite{asheghi1998temperature, ju1999phonon, ju2005phonon, hao2006thermal, liu2006thermal} and thermoreflectance techniques.\cite{aubain2010determination, aubain2011plane} The suspended films (open markers) were measured using suspended island,\cite{tang2010holey, yu2010reduction} Raman thermometry,\cite{chavez2014reduction, neogi2015tuning} and transient thermal grating (TTG) techniques.\cite{cuffe2015reconstructing} The predictions are from an analytical model,\cite{mcgaughey2011size} the Matthiessen rule (MR),\cite{jain2013phonon, wang2014computational} free path sampling (FPS), \cite{jain2013phonon} and the discrete ordinate method (DOM).\cite{fu2017electron} Predictions using the Fuchs-Sondheimer (FS) model\cite{cuffe2015reconstructing} were made with bulk phonon mean free paths predicted by Esfarjani et al.\cite{Esfarjani2011Heat} (FS + Esfarjani et al.) or Holland\cite{Holland1963Analysis} (FS + Holland). The experimental bulk thermal conductivity\cite{inyushkin2004isotope} is represented by the dashed horizontal line.}
\label{F-thermalconductivitydata}
\end{figure}

Extensive experimental measurements have been performed on films where phonons dominate thermal transport, demonstrating the significant in-plane thermal conductivity reduction as the film thickness is reduced. A collection of data for silicon films at a temperature of 300 K with thicknesses between 9 nm and 1.5 $\mu$m is plotted in Fig.~\ref{F-thermalconductivitydata}. The measurements were performed using $\rm 3-\omega$,\cite{asheghi1998temperature, ju1999phonon, ju2005phonon, liu2006thermal, hao2006thermal} thermoreflectance (TR),\cite{aubain2010determination, aubain2011plane} suspended island,\cite{tang2010holey, yu2010reduction} Raman thermometry,\cite{chavez2014reduction,neogi2015tuning} and transient thermal grating (TTG)\cite{cuffe2015reconstructing} techniques. For the $\rm 3-\omega$ and TR measurements, the films were embedded between a metal capping layer and a substrate. For the suspended island, Raman thermometry, and TTG measurements, the films were suspended. While the thickness-dependent trend is clear, the thermal conductivities at any given thickness are scattered, with a spread of up to 48 W/m-K at a thickness of 150 nm, where the average value is 72 W/m-K.
The same thermal conductivity trend, i.e., a reduction with decreasing film thickness, has been measured for GaN,\cite{beechem2016size} black phosphorus,\cite{luo2015anisotropic} and h-BN.\cite{jo2013thermal}

\begin{figure}[tb]
\centering
\includegraphics{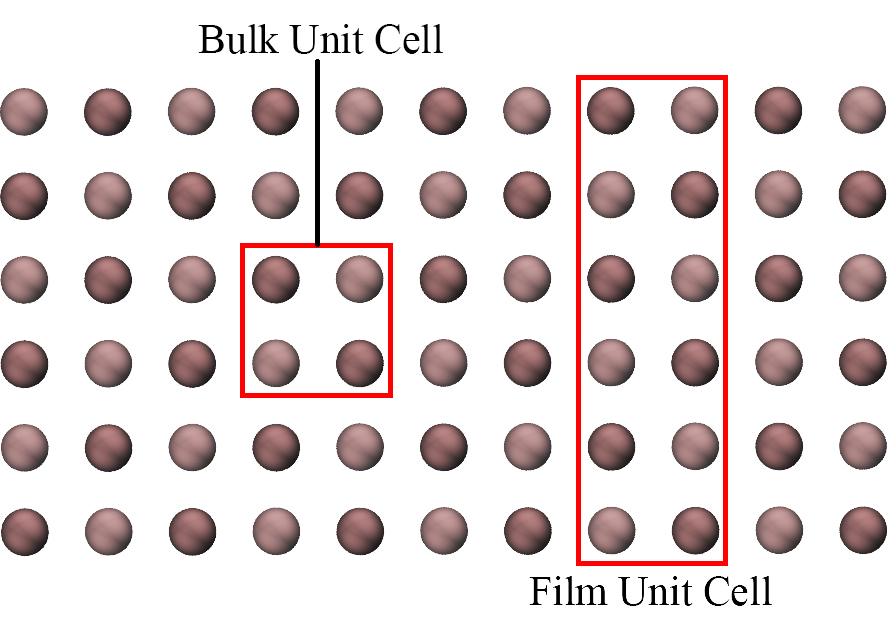} 
\caption{Bulk and film unit cells in a three unit cell thick unrelaxed LJ argon film. The conventional cell with four atoms is used as the basis in bulk and to build the film unit cell.}
\label{F-unitcell}
\end{figure}

The origin of the thermal conductivity reduction in a film can be probed theoretically. Phonon-boundary scattering reduces the bulk mean free paths (MFPs) and thus reduces thermal conductivity. The film MFP for a given phonon mode can be obtained by combining the intrinsic bulk phonon-phonon scattering MFP with a boundary scattering model using the Matthiessen rule\cite{mcgaughey2011size,jain2013phonon,wang2014computational} or the Fuchs-Sondheimer (FS) suppression function.\cite{cuffe2015reconstructing}
Phonon-boundary scattering can also be simulated using numerical techniques,\cite{mcgaughey2012nanostructure, yang2004thermal} which are useful for arbitrary geometries such as nanoporous films and nanocomposites.\cite{jain2013phonon,tang2013thermal,fu2014thermal,fu2017electron} All of these methods use bulk phonon properties as the starting point. The phonon transport is thus three-dimensional (3D) and is based on the bulk unit cell shown in Fig.~\ref{F-unitcell}, which has a corresponding 3D Brillouin zone. While the focus here is on in-plane thermal transport, similar tools can be applied to investigate cross-plane thermal transport.\cite{sellan2010cross, vermeersch2016cross}

The results from a series of in-plane thermal conductivity predictions based on the 3D treatment for silicon are plotted in Fig.~\ref{F-thermalconductivitydata}.\cite{mcgaughey2011size, jain2013phonon, wang2014computational,fu2017electron, cuffe2015reconstructing} The bulk phonon properties were obtained from lattice dynamics calculations and a solution of the Boltzmann transport equation (BTE), which we will refer to as 3D-BTE, under the relaxation time approximation.\cite{mcgaughey2011size, jain2013phonon, wang2014computational,fu2017electron, cuffe2015reconstructing} 
In the lattice dynamics calculations, the atomic interactions were modeled using an empirical model\cite{mcgaughey2011size, cuffe2015reconstructing, Holland1963Analysis} or density functional theory calculations.\cite{jain2013phonon, wang2014computational, fu2017electron, cuffe2015reconstructing, Esfarjani2011Heat} All calculations were performed on suspended films with an assumption of fully-diffuse phonon-boundary scattering. As with the experimental measurements, the thermal conductivity predictions at any given film thickness are scattered, with a spread of up to 27 W/m-K at a thickness of 1 $\rm \mu$m, where the average value is 107 W/m-K.

The agreement between the measurements and predictions plotted in Fig.~\ref{F-thermalconductivitydata} worsens as the film thickness gets smaller. Differences may originate from: 
(i) Some of the experimental films being embedded, while all the predictions are for suspended films. For an embedded film, there is the potential for film phonons to transmit into the surrounding materials and vice-versa, while for a suspended film, all phonon-boundary interactions lead a specular or diffuse reflection. 
(ii) The presence of a native oxide layer on the experimental films.\cite{neogi2015tuning} While thermal transport in the oxide layer is not considered explicitly in the predictions, a predetermined parameter (i.e., the specularity) can be used to model the surface roughness. 
(iii) The validity of the theoretical framework.  Our objective here is to examine (iii). Specifically, we will probe the assumption of using phonon modes based on the 3D bulk description.

The rigorous description of the phonons modes in a film is based on a unit cell that spans the film thickness and has a corresponding two-dimensional (2D) Brillouin zone, as shown in Fig.~\ref{F-unitcell}. Phonons only advect in the in-plane directions and the concepts of cross-plane transport and boundary scattering are not meaningful. For thick enough films, as shown in Fig.~\ref{F-thermalconductivitydata}, the phonon transport can be adequately described using the 3D Brillouin zone and a boundary scattering model. But for thin enough films, which we will refer to as ultrathin films, the 3D treatment breaks down. This phenomenon is known as phonon confinement.\cite{turney2010plane, balandin1998significant} The identification of the phonon confinement regime is a central objective of this work.

Kim et al.\cite{Kim2018mapping} recently proposed an algorithm to map between 2D and 3D phonon modes, which they applied to Lennard-Jones (LJ) argon and multi-layer graphene films. A near one-to-one correspondence was discovered for thicknesses greater than 5 nm for LJ argon and for all thicknesses of graphene. Their study only considered harmonic phonon properties, however, and not the impact of confinement on phonon scattering and thermal conductivity. 
2D harmonic lattice dynamics calculations have also been applied to predict the phonon properties of silicon films with surface nanopillars.\cite{Davis:2014d32, honarvar2018two} 

Beyond harmonic lattice dynamics, the 2D treatment is commonly applied to model phonon transport and scattering in 2D materials such as graphene,\cite{lindsay2010flexural} multi-layer graphene,\cite{lindsay2011flexural} Mo$\rm S_2$,\cite{li2013thermal} and black phosphorene\cite{jain2015strongly} using anharmonic lattice dynamics calculations and the BTE. While the 2D treatment is rigorous, the computational cost of the BTE-based method, which scales with the number of atoms in the unit cell to the fourth power, makes calculations on films challenging. To our knowledge, the 2D treatment at the anharmonic level has not been applied to films built from common semiconductors such as silicon or GaN. 

Instead, previous researchers investigated confinement using the bulk 3D phonon modes. In the phonon depletion approach, only bulk phonon modes with cross-plane wavelengths that fit inside the film are considered.\cite{turney2010plane,wang2014computational,lindsay2010diameter} I.e., the finite sampling of the 3D Brillouin zone in the cross-plane direction is set by the film thickness. Phonon-boundary scattering must still be included.
Turney et al.\cite{turney2010plane} used the phonon depletion approach in 3D-BTE and the Green-Kubo (GK) method in MD simulations to predict the thermal conductivity of Lennard-Jones (LJ) argon films with thickness as small as 1 nm and Stillinger-Weber silicon films with thickness as small as 17.4 nm. Wang and Huang\cite{wang2014computational} also investigated in-plane thermal transport in silicon films using 3D-BTE and phonon depletion with input from density functional theory calculations. 

Herein, we perform 2D lattice dynamics calculations combined with an iterative solution to the BTE (2D-BTE) to study phonon transport in suspended films of LJ argon (a soft material) and Tersoff silicon (a stiff material) with thickness as small as two conventional unit cells (i.e., four atomic layers for argon and eight atomic layers for silicon). The 2D-BTE approach uses a unit cell that spans the film thickness (Fig.~\ref{F-unitcell}), thus directly treating the phonon transport as 2D without any phonon-boundary scattering. In doing so, we remove all approximations associated with the finite cross-plane dimension used in previous 3D-BTE studies. 
MD simulations are performed to obtain the finite temperature atomic structures and to predict thermal conductivity benchmarks. We also perform 3D-BTE calculations with the Fuchs-Sondheimer suppression function (3D-BTE-FS) and phonon-depletion to assess their validity. This multi-method approach allows us to unravel the similarities and differences between phonon transport in corresponding 2D film and 3D bulk systems. We use empirical potentials to describe the atomic interactions due to the large computational costs associated with the large unit cells.  Our findings will be relevant to other film systems.

The rest of the paper is organized as follows. 
The potential parameters, brief descriptions of the MD and BTE methods, and computational details are provided in Sec.~\ref{sec:meth}. Results and discussion for the LJ argon and silicon films are presented in Secs.~\ref{sec:argon} and~\ref{sec:silicon}. Included are descriptions of the finite temperature atomic structures, anharmonicity analysis performed using root-mean-squared (RMS) displacements, and phonon properties and thermal conductivities predicted from the 2D-BTE, 3D-BTE, and MD treatments. Of note are predictions of unusual thermal conductivity trends for very thin films, underestimations of thermal conductivity from the 3D-BTE-FS description, and incorrect trends predicted by the depletion approach. The results are interpreted using the density of states (DOS), mode lifetimes, and thermal conductivity accumulation functions. The findings are summarized in Sec.~\ref{sec:sum}.  

\clearpage

\section{\label{sec:meth}Methodology}

\subsection{Interatomic Potentials and Structures}

Solid argon is a face-centered cubic crystal with a one-atom basis. The potential energy $\phi$ between two atoms $i$ and $j$ is modeled using the two-body LJ potential\cite{Ashcroft1976solid}
\begin{eqnarray}
\phi(r_{ij}) = 4\epsilon_{\rm LJ}\left[ \left( \frac{\sigma_{\rm LJ}}{r_{ij}} \right)^{12} - \left( \frac{\sigma_{\rm LJ}}{r_{ij}} \right)^{6}  \right]. \label{LJ}
\end{eqnarray}
Here, $r_{ij}$ is the distance between the two atoms, $\epsilon_{\rm LJ}$ is the energy scale ($1.67 \times 10^{-21}$ J), and $\sigma_{\rm LJ}$ is the length scale ($3.4 \times 10^{-10}$ m). 
The atomic mass is $6.63 \times 10^{-26}$ kg. The potential energy is shifted at a cut-off radius of $3\sigma_{\rm LJ}$. We found that using the typical cutoff radius of $2.5\sigma_{\rm LJ}$ led to imaginary modes in the film phonon dispersion curves that disappeared when the cutoff radius was increased to $3\sigma_{\rm LJ}$. 
Temperatures of 5, 10, and 20 K are considered. These low temperatures (the melting temperature of LJ argon is around 90 K) allow for comparisons with the BTE predictions, which include up to three-phonon scattering events. The significant anharmonicity in LJ argon at higher temperatures leads to non-negligible four-phonon scattering, which is not considered here.\cite{turney2009predicting,feng2016quantum}

Silicon has a diamond crystal structure and is modeled using the three-body Tersoff potential.\cite{tersoff1989modeling} All calculations are performed at a temperature of 200 K to limit the effects of four-phonon scattering.\cite{feng2016quantum} We do not consider any reconstruction of the film surfaces. All structures are stable in the MD simulations and exhibit only real frequencies in their phonon dispersions.

The films are initialized as supercells built from $N_x \times N_y \times N_z$ bulk conventional unit cells (i.e., a simple cubic lattice with a four-atom basis for argon and an eight-atom basis for silicon). An in-plane size of $N_x = N_y = 8$ is used, while $N_z$ is the number of unit cells in the cross-plane direction. Periodic boundary conditions are applied in all three directions. A vacuum region of three conventional units cells, which is larger than the range of the atomic interactions, is placed between the top and bottom surfaces of the film to ensure that the film is isolated and suspended.

Computational costs limit the film thicknesses that can be explored, even with the use of empirical potentials.
For an efficient MD simulation, the computational cost scales linearly with the number of atoms in the computational cell. 
For the 2D-BTE and 3D-BTE calculations, however, the computational cost scales as {$\mathcal{O}(n_{\rm unitcell}^4M^2)$}, where $n_{\rm unitcell}$ is the number of atoms in the unit cell and $M$ is the number of wave vectors in the first Brillouin zone.\cite{turney2009predicting} 
As a result, although we can perform MD simulations for films with thicknesses up to 100 unit cells for argon (25,600 atoms) and 40 unit cells for silicon (20,480 atoms), the 2D-BTE calculations are limited to thicknesses of 20 unit cells for argon and 8 unit cells for silicon.

\subsection{Molecular Dynamics Simulation\label{sec:MD}}

The MD simulations on the bulk and film structures are performed using the open-source LAMMPS\cite{plimpton1995fast} and GPUMD\cite{fan2017efficient} packages.
LAMMPS is applied to obtain the finite temperature atomic structures and the RMS displacements for both argon and silicon and to predict the thermal conductivity of argon using the Green-Kubo method. 
For the silicon thermal conductivity predictions, GPUMD is used to calculate the heat current vector required in the Green-Kubo method because it provides the correct formulation for the Tersoff three-body potential.\cite{fan2015force}
The time-step is 2 fs for both argon and silicon.

To obtain the zero-pressure bulk atomic structures, an MD simulation is run on an $8 \times 8 \times 8$ system in the $NVT$ ensemble (a constant number of atoms, volume, and temperature) for $5\times10^5$ time steps to set the temperature. 
The system is then evolved in the $NPT$ ensemble (a constant number of atoms, pressure, and temperature) at zero pressure and the set temperature for $2\times10^6$ time steps. The zero-pressure lattice constant is obtained by time-averaging the simulation cell size in the $x$-, $y$-, and $z$- directions and dividing by $N_x=N_y=N_z=8$. The simulation cell size is collected every 10 time steps and the averaging is performed over the last $10^6$ time steps in the $NPT$ ensemble.

For each film thickness and temperature, the zero-pressure in-plane lattice constant is determined  using the same procedure as for bulk, but by time-averaging the simulation cell size in the $x$- and $y$- directions.
Fixing the in-plane lattice constant at this value, an $NVT$ simulation is run for $2\times10^6$ time steps to allow the film to relax in the cross-plane direction. 
Every ten time steps, the average $z$-location of each atomic layer is obtained by averaging the $z$-positions of all the atoms in that layer. The results are then time-averaged over the last $10^6$ time steps. 
The resulting atomic structure is used as the starting point for all further MD simulations and the 2D-BTE calculations.
The film thickness is defined as the product of the number of atomic layers and the average layer separation in the cross-plane direction.

Thermal conductivity is predicted using the Green-Kubo method, which is an equilibrium MD technique based on the fluctuation-dissipation theorem.\cite{McQuarrie2000statistical} 
The system is first equilibrated in the $NVT$ ensemble for $2\times10^6$ time steps and then in the $NVE$ ensemble for $2.5\times10^6$ time steps. 
The heat current vector is then recorded every five time steps in the $NVE$ ensemble for $2.5\times10^7$ time steps for argon and $2.5\times10^8$ time steps for silicon and then used to obtain the heat current autocorrelation function (HCACF). The correlation time is $2.5\times10^5$ time steps for argon and $2.5\times10^6$ time steps for silicon.
The HCACFs for thirty seeds with randomized initial velocities are collected to sufficiently explore the phase space. 
The thermal conductivity is obtained by integrating the HCACF for each seed and then averaging over the last 25\% of the correlation time. 
We determined the uncertainty by calculating the 95\% confidence interval, $1.96\frac{\sigma}{\sqrt{n}}$, where $\sigma$ and $n$ are the standard deviation and number of seeds. The maximum uncertainty for both material systems for all film thicknesses and temperatures is 7\%. In terms of size effects, an in-plane size of $N_x = N_y = 8$ provides a thermal conductivity that is converged to within 5\%.

\subsection{Lattice Dynamics Calculations}

All lattice dynamics and BTE calculations are performed using in-house codes.\cite{mcgaughey2019phonon, Jainphdthesis} From the linearized BTE and the Fourier law, the thermal conductivity in direction $\alpha$, $k_{\alpha}$, can be expressed as
\begin{eqnarray}
k_\alpha = \sum_{\pmb{\kappa},\nu} c_\mathrm{ph} \kv v_{\rm g,\alpha}^2 \kv \tau_\alpha \kv, \label{E-kBTE}
\end{eqnarray}
where a phonon mode is identified by its wave vector $\pmb{\kappa}$ and a polarization branch $\nu$, $c$ is the volumetric specific heat, and $v_{\rm g,\alpha}$ is the $\alpha$-component of the group velocity vector, which can be obtained from the phonon dispersion. 
The phonon dispersion (i.e., the relationship between the frequencies and wave vectors) is calculated from harmonic lattice dynamics calculations under the quasi-harmonic approximation (i.e., using the finite-temperature structures, as described in Sec.~\ref{sec:MD}). 
The phonon lifetime, $\tau_\alpha$, corresponds to a heat flow in the $\alpha$-direction. Scattering due to three-phonon interactions is calculated using anharmonic lattice dynamics calculations combined with an iterative solution to the linearized BTE.\cite{Wallace1972Thermodynamics, Reissland2004quantum, Wardphdthesis, Jainphdthesis, mcgaughey2019phonon} Classical (i.e. Maxwell-Boltzmann) statistics are used to allow for comparison with the inherently classical MD simulations, such that $c_{\rm v} = k_{\mathrm B}/V$, where $k_{\mathrm B}$ is the Boltzmann constants and $V$ is the simulation cell volume. We do not consider any additional scattering mechanisms for the bulk systems or for films modeled using the 2D Brillouin zone (i.e., a unit cell that spans the film thickness, as shown in Fig.~\ref{F-unitcell}). 

For films modeled using the bulk 3D Brillouin zone, we account for phonon-boundary scattering using the Fuchs-Sondheimer model. The lifetime is calculated by modifying the intrinsic lifetime $\tau_{\rm pp, \alpha}$ with\cite{turney2010plane}
\begin{eqnarray}
\tau_\alpha \kv= F\kv \tau_{\rm pp, \alpha} \kv, 
\end{eqnarray}
where $F$ is a mode-dependent scaling factor (i.e., a suppression function) given by 
\begin{eqnarray}
\label{eq_fs}
F\kv = 1 - \frac{1-p\kv}{\delta\kv}\frac{1-\rm exp[-\delta\kv]}{1-p\kv \rm exp[-\delta\kv]}.
\end{eqnarray}
Here, $p \kv$ is the mode-dependent specularity parameter, which ranges from zero for completely diffuse scattering to unity for completely specular scattering. 
Unless noted, we set $p \kv$ to be zero. $\delta \kv$ is equal to $L_{z}/(|v_{{\rm g},z}|\tau_{\rm pp, \alpha})$ where $v_{{\rm g},z}$ is the cross-plane ($z$) component of the bulk group velocity vector and $L_{z}$ is the film thickness. The scaling factor $F\kv$ decreases monotonically with decreasing film thickness, thus leading to a lower thermal conductivity.

The harmonic (second-order) and cubic (third-order) force constants required in the 2D/3D-BTE calculations are obtained by applying displacements of 0.005 $\angstrom$ to selected atoms and performing finite differencing of the resulting forces.\cite{mcgaughey2019phonon} For argon, the harmonic (cubic) force constants are extracted by applying a cutoff radius to include up to the seventh (first) nearest-neighbors [1.02 nm (0.58 nm)].
For silicon, the harmonic cutoff radius is set as the potential cutoff (0.3 nm) and the anharmonic cutoff radius includes up to the third nearest-neighbor (0.43 nm).
Phonon properties and thermal conductivity are calculated on a $24 \times 24 \times 24$ wave vector grid for bulk and a $24 \times 24$ wave vector grid for the films. Further increasing the resolution changes the predicted thermal conductivities by less than 1\%. 

\clearpage
\section{\label{sec:argon} Lennard-Jones Argon Results}
\subsection{\label{sec:argonstr}Film Atomic Structure}

\begin{figure}[b]
\centering
\includegraphics{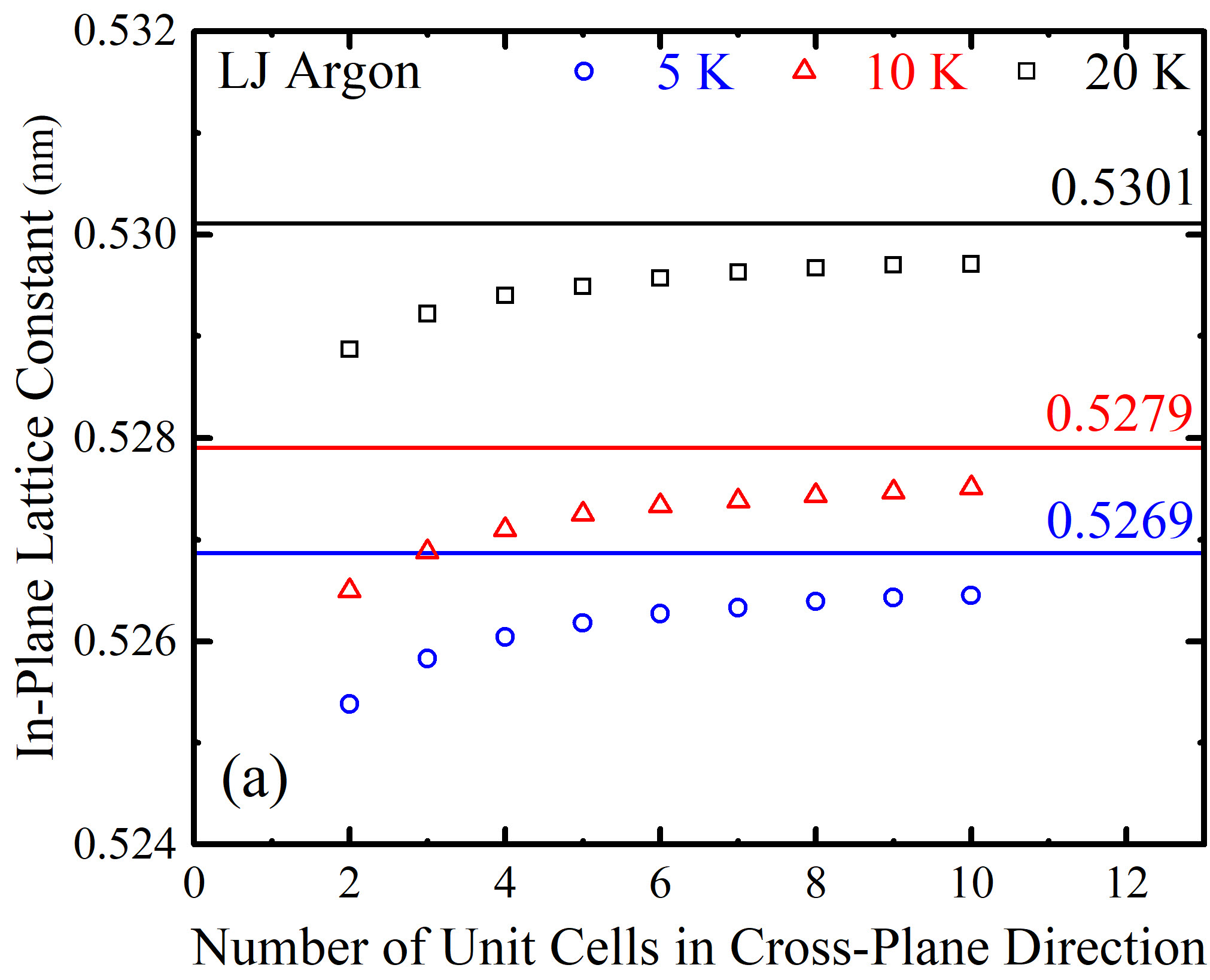}\\
\vspace{.1in} 
\includegraphics{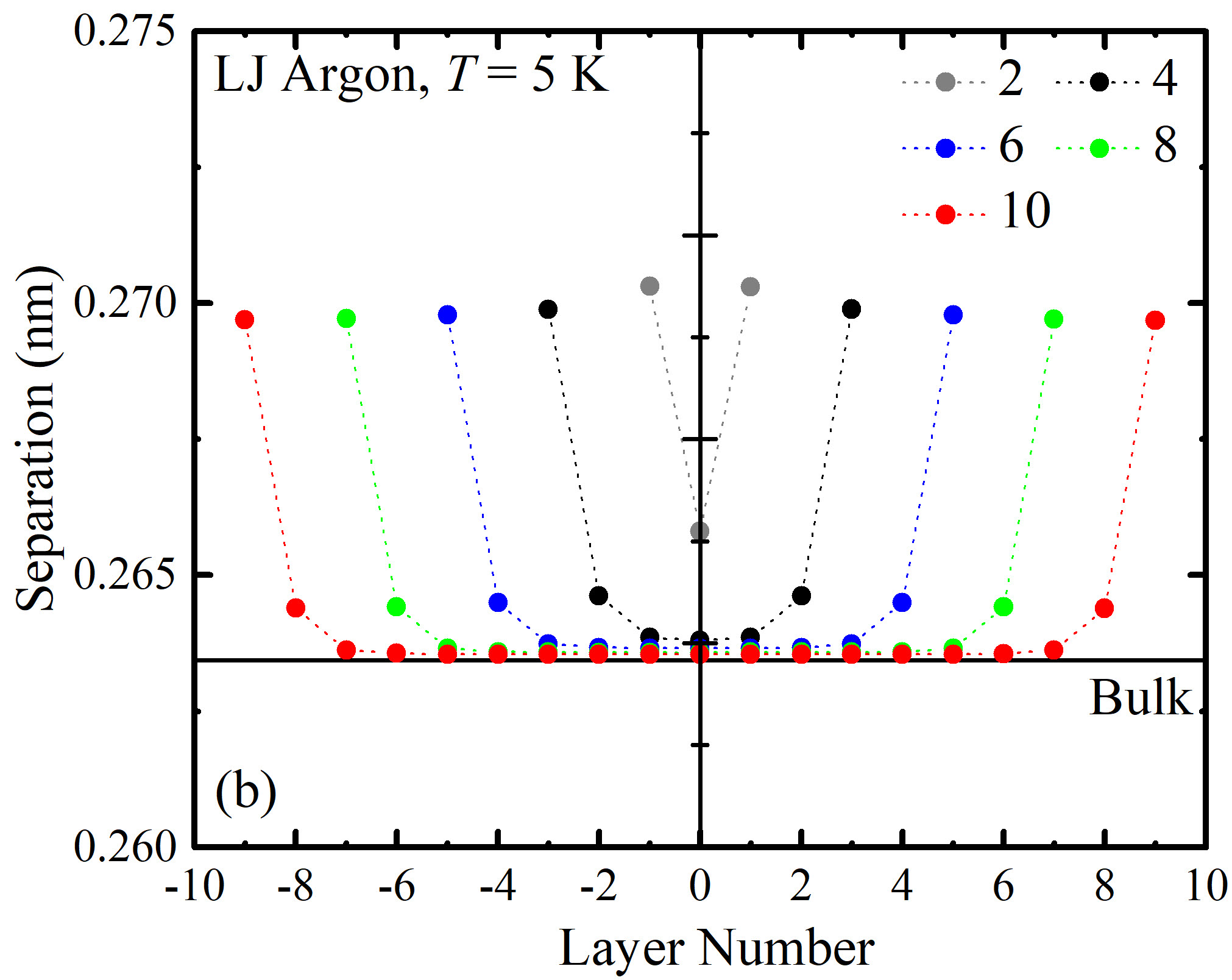}
\caption{(a) In-plane lattice constant versus film thickness for argon films at temperatures of 5 (blue), 10 (red), and 20 (black) K. The solid horizontal lines represent the bulk lattice constants. 
(b) Separations between atomic layers in the cross-plane direction versus the layer number, which represents the location in the film. 
For example, for the two unit cell film, there are four layers and three layer separations. ``0" corresponds to the separation between the two central layers while ``-1" and ``1" correspond to the separations involving the surface layers. The solid horizontal line represents the bulk layer separation. The numbers in the legend correspond to the film thickness in conventional unit cells.}
\label{F-structure_argon}
\end{figure}

The in-plane lattice constants of the argon films with thicknesses between two and ten unit cells (1.1 - 5.3 nm) at temperatures of 5, 10, and 20 K are plotted in Fig.~\ref{F-structure_argon}(a). Also plotted are the corresponding bulk lattice constants, which are provided in Table \ref{t-argon}. At all temperatures, the in-plane film lattice constant increases monotonically with increasing thickness and approaches the bulk value.

\begin {table}[t]
\caption {Lattice constants, linear RMS displacements along one of the Cartesian directions, and thermal conductivities ($k_{\rm MD-GK}$ and $k_{\rm 3D-BTE}$) for bulk argon at temperatures of 5, 10, and 20 K and bulk silicon at a temperature of 200 K. The 3D-BTE results are converged to within 1\% based on the wave vector resolution.} \label{t-argon} 
\begin{center}
\begin{tabular}{ c | c | c | c| c }
\hline
\hline
System & \multicolumn{3}{c|}{Argon} & Silicon \\
\hline
Temperature & 5 K & 10 K & 20 K & 200 K\\
\hline
Lattice Constant (nm) & 0.5269 & 0.5279 & 0.5301 & 0.5440 \\
Linear RMS Displacement (nm) & 0.0048 & 0.0069 & 0.0101 & 0.0054 \\
$k_{\rm MD-GK}$ (W/m-K) & $8.9 \pm 0.2$ & $3.9 \pm 0.2$ & $1.5 \pm 0.1$ & $414 \pm 7$ \\
$k_{\rm 3D-BTE}$ (W/m-K) & 9.0 & 4.1 & 1.7 & 464 \\
\hline
\hline
\end{tabular}
\end{center}
\end{table}

The cross-plane atomic layer separations for films with thicknesses between two and ten unit cells at a temperature of 5 K are plotted in Fig.~\ref{F-structure_argon}(b). Note that one unit cell is made up of two layers. Also plotted is the bulk layer separation, which corresponds to half of the bulk lattice constant. The layer separation varies across the film thickness and, as expected, is symmetrical about the centerline. The outermost layer separation for all films is larger than that in middle due to the weak bonding of the LJ potential. The layer separations inside the films are close to the bulk value.
The exception is the two unit cell film, where the two middle layers are directly adjacent to the surface layers. The deviation from the bulk layer separation for this film at the centerline is 1\%, which is large compared to the 0.04\% deviation for the 10 unit cell film. Due to the cross-plane expansion and the in-plane contraction, the resulting film structures are anisotropic, an effect that becomes stronger as the thickness decreases. 
Similar results were found at temperatures of 10 and 20 K, as shown in Figs. S1(a) and S1(b) of the Supplemental Material.\cite{SM} The anisotropy suggests that the anharmonicity present in the films will be different from that in bulk, which we explore next in Sec.~\ref{sec:argonanhar}.

\subsection{\label{sec:argonanhar}Anharmonicity}

Phonon-phonon scattering is a result of anharmonicity and is the origin of the finite thermal conductivities of our bulk structures modeled using MD-GK and 3D-BTE and our film structures modeled using 2D-BTE.
Before presenting the thermal conductivity predictions in Sec.~\ref{sec:argonk}, we first examine the anharmonicity by considering the atomic RMS displacements.\cite{parrish2014origins} As an atom deviates further from its equilibrium position, it is exposed to greater anharmonicity. Increasing anharmonicity is typically associated with increasing temperature, but it can also be a result of a free surface.

\begin{figure}[tb]
\centering
\includegraphics{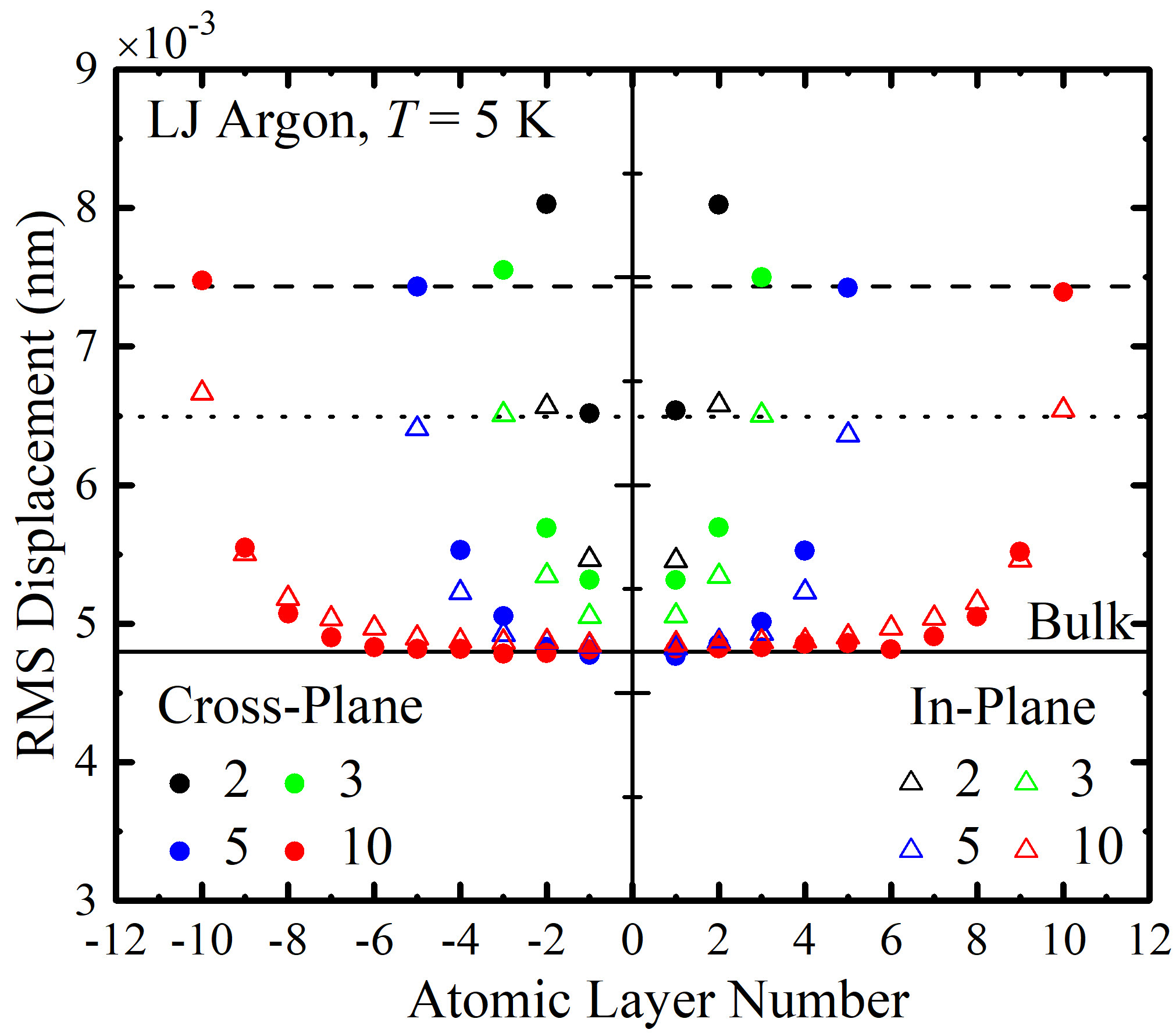}
\caption{RMS displacement in the cross-plane and in-plane directions versus atomic layer number for argon at a temperature of 5 K. The solid line represents the bulk RMS displacement along one Cartesian direction. The dashed and dotted lines represent the values that the surface atoms approach in cross-plane and in-plane directions. The numbers in the legend correspond to the film thickness in conventional unit cells.}
\label{F-rms_all_argon}
\end{figure}

The RMS displacements are calculated from MD simulations in the $NVT$ ensemble. Data were collected every ten time steps from a $2\times10^6$ time step simulation and averaged over each layer.
The in-plane and cross-plane RMS displacements for film thicknesses between two and ten unit cells at a temperature of 5 K are plotted in Fig.~\ref{F-rms_all_argon}. The in-plane RMS displacement is averaged over the $x$- and $y$- directions. Also plotted is the bulk value along one of the Cartesian directions, which is provided in Table \ref{t-argon}. The in-plane and cross-plane RMS displacements increase in moving from the centerline to the surfaces.
At the surfaces, the cross-plane RMS displacement is larger than the in-plane value, which is a result of the structural anisotropy discussed in Sec.~\ref{sec:argonstr}. 

From Fig.~\ref{F-structure_argon}(b), for film thicknesses greater than two unit cells, the two outermost layers have larger-than-bulk separations while the other layers are bulk-like. For the RMS displacements, we see that the effect of the surfaces penetrates three layers deep.
The two unit cell film has the largest RMS displacements as it contains no bulk-like atoms.
The larger displacements of the surface atoms will lead to larger anharmonicity in that part of the film.  As the film thickness increases above five unit cells, the surface atom RMS displacements converge, indicated by the dashed and dotted lines for cross-plane and in-plane in Fig.~\ref{F-rms_all_argon}.
The converged in-plane and cross-plane values are 0.0065 nm and 0.0074 nm, which are 35\% and 54\% larger than the bulk value. 
Similar results are observed at temperatures of 10 and 20 K, as shown in Figs. S2(a) and S2(b).

\clearpage

\subsection{\label{sec:argonk}Thermal Conductivity}

\begin{figure}[b]
\centering
\includegraphics{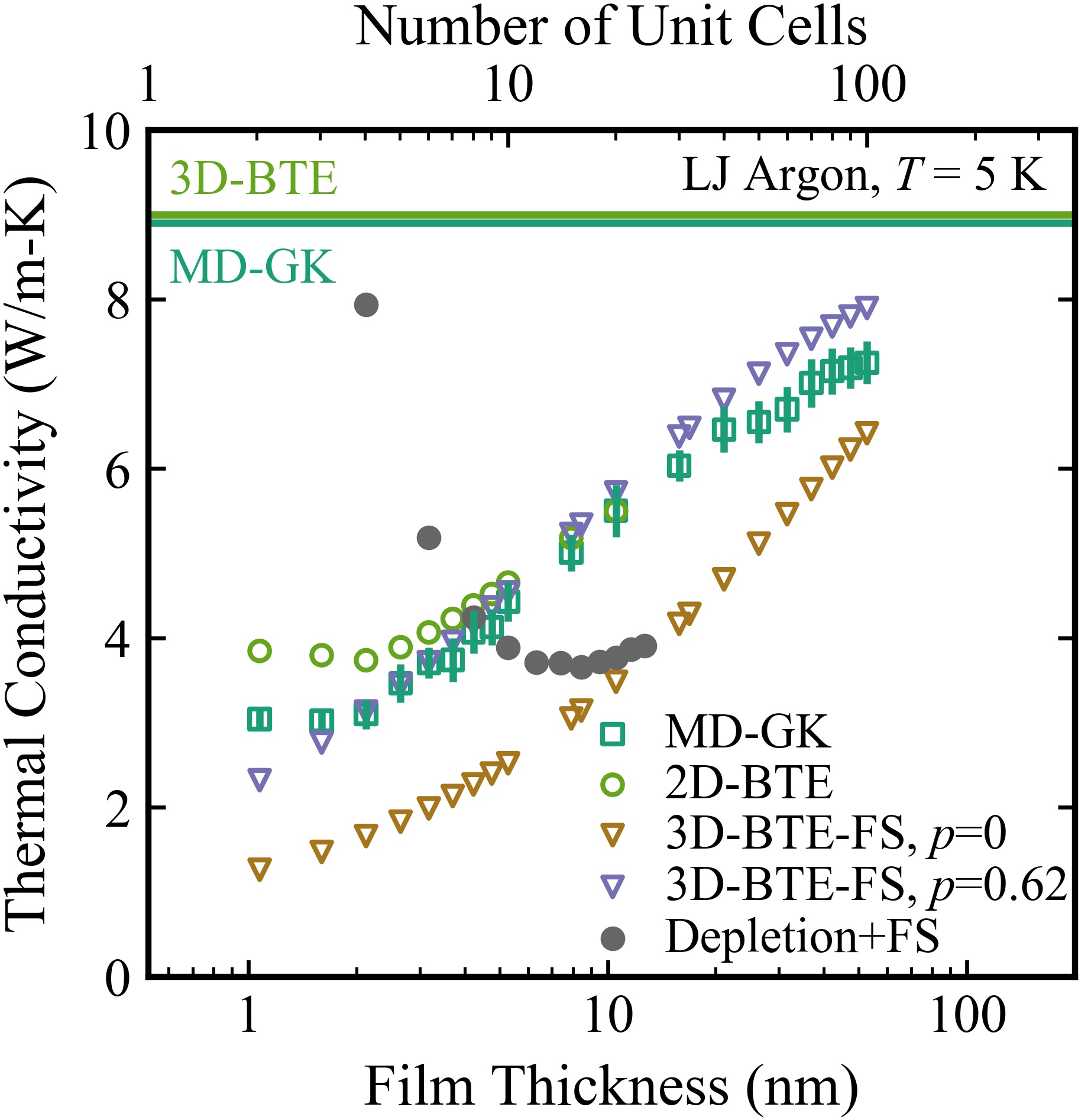}\\
\vspace{0.1in}
\includegraphics{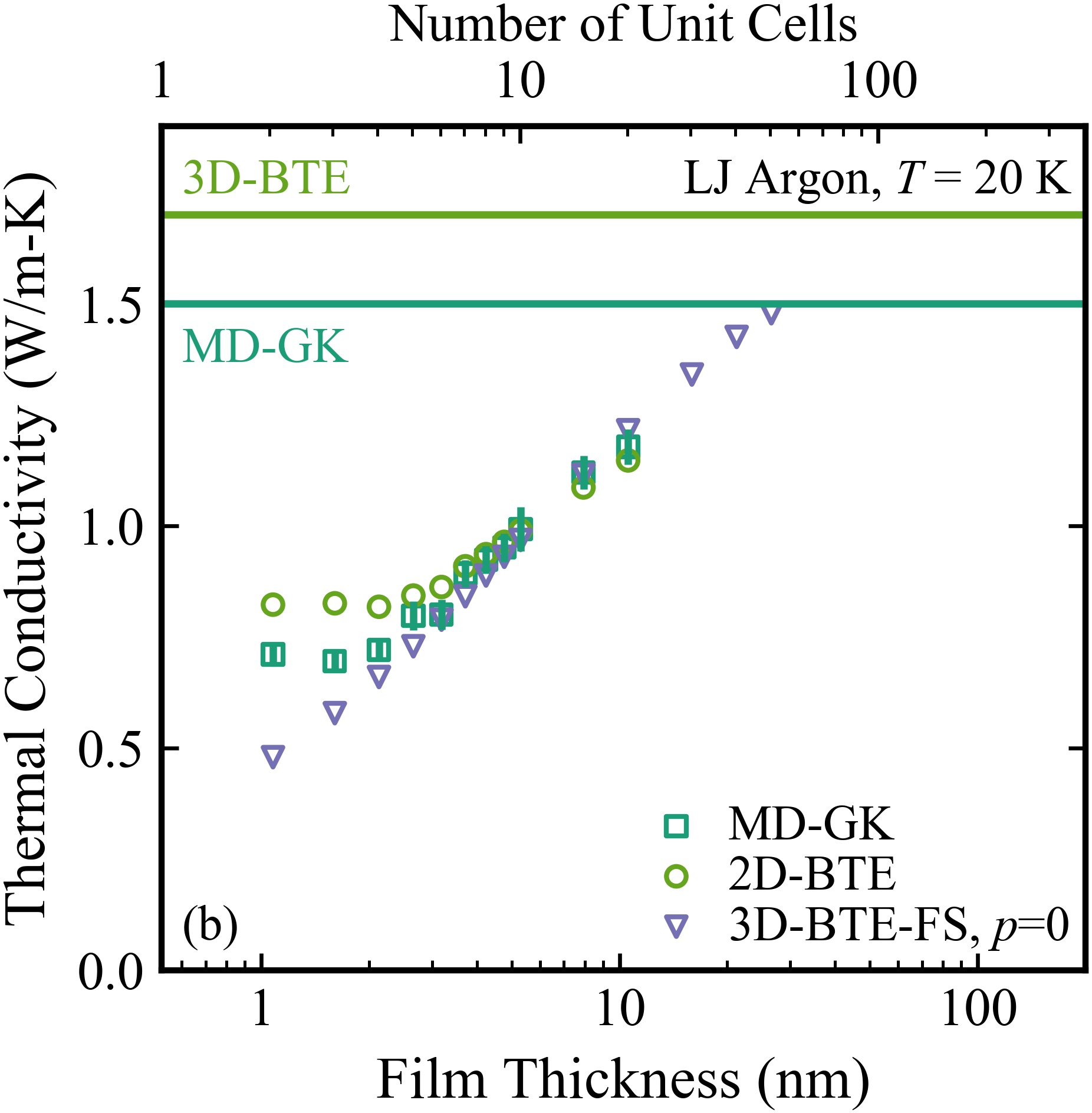}
\caption{Thermal conductivity versus thickness for argon films at temperatures of (a) 5 K and (b) 20 K. 
The results from the MD-GK, 2D-BTE, and 3D-BTE (with bulk or depleted phonons combined with the Fuchs-Sondheimer boundary scattering model) methods are presented. The Fuchs-Sondheimer boundary scattering model is applied with specularity parameters $p$ of 0 (both temperatures) and 0.62 (5 K). The horizontal solid lines indicate the bulk thermal conductivities from 3D-BTE and MD-GK.}
\label{F-thermalconductivity_argon}
\end{figure}

The thermal conductivities of bulk argon predicted from the MD-GK and 3D-BTE methods at temperatures of 5, 10, and 20 K are provided in Table~\ref{t-argon}. The two methods can be compared based on how they treat the anharmonicity. MD simulations include the full anharmonicity and thus all orders of phonon-phonon interactions. We thus take the MD-GK thermal conductivities as a benchmark. The 3D-BTE and 2D-BTE approaches, as implemented, only consider up to three-phonon processes. When considering equivalent structures, potential parameters, and temperatures, the thermal conductivity prediction from 3D-BTE or 2D-BTE should therefore be larger than that from MD-GK, with a difference that increases with increasing temperature. The data in Table~\ref{t-argon} follow this trend. Of note is the agreement at a temperature of 5 K, indicating that three-phonon processes are sufficient to capture phonon transport in the bulk phase.

The in-plane thermal conductivity variation with film thickness for argon films predicted using the 2D-BTE and MD-GK methods at a temperature of 5 K is plotted in Fig.~\ref{F-thermalconductivity_argon}(a). The bulk thermal conductivities from Table~\ref{t-argon} are included as horizontal lines.
Both methods predict a decrease in thermal conductivity with decreasing thickness up to a thickness of four unit cells ($2.1$ nm). Below that thickness, the thermal conductivity plateaus. The origin of the plateau is investigated in Sec.~\ref{sec:argonplateau} by analyzing the mode-level phonon properties.
The 2D-BTE film thermal conductivities are always higher than the MD-GK values, with a difference that increases with decreasing thickness. 
The increase of the difference is a result of the increased anharmonicity at the film surfaces, which becomes more prominent as the thickness decreases, as discussed in Sec.~\ref{sec:argonanhar}. The 2D-BTE calculations, which are based on small deviations from the equilibrium positions, do not capture the increased anharmonicity and thus over-predict the thermal conductivity for the smallest thicknesses. 
In using the MD-GK method to model LJ argon films at a temperature of 20 K, Turney et al.\cite{turney2010plane} observed a monotonically decreasing thermal conductivity down to thicknesses as small as two unit cells. 
One possible origin of this different trend compared to our result is that Turney et al. used a fixed in-plane lattice constant equal to the bulk value,\cite{replicate} while we relaxed the film structures to zero-pressure as described in Sec.~\ref{sec:MD}. 

Film thermal conductivities predicted from 3D-BTE combined with the Fuchs-Sondheimer phonon-boundary scattering model are also plotted in Fig.~\ref{F-thermalconductivity_argon}(a). Three data sets are shown.
In the first data set, denoted as 3D-BTE-FS ($p=0$) and plotted as inverted brown triangles, the full set of phonon modes from the Brillouin zone are used and the phonon-boundary scattering is taken as completely diffuse. For the two unit cell film, the thermal conductivity from 3D-BTE-FS ($p=0$) is 68\% smaller than that from the rigorous 2D-BTE treatment. The underestimation decreases as the film thickness increases and its modes become more bulk-like. In the second data set, denoted as 3D-BTE-FS ($p=0.62$) and plotted as inverted purple triangles, the specularity parameter $p$ is increased to 0.62 to match the 2D-BTE data for thicknesses greater than ten unit cells. With this choice, the plateau at very small thicknesses is not observed. We note that while the specularity parameter is in general phonon mode-dependent, we consider a mode-independent value to examine its overall impact on the predictions.

In the third data set, denoted as Depletion+FS and plotted as filled grey circles, we consider the phonon depletion approach discussed in Sec.~\ref{sec:intro}\cite{turney2010plane,wang2014computational} with completely diffuse surfaces. The phonon wave vectors in the cross-plane direction are limited to discrete values by the constraint
\begin{eqnarray}
\kappa_z = \frac{2\pi l_z}{aN_z},
\label{E-depletion}
\end{eqnarray}
where $l_z$ is an integer with a magnitude less than $N_z/2$ and $a$ is the bulk lattice constant. 
The phonon properties are then calculated using the reduced list of wave vectors and the Fuchs-Sondheimer model. 
For the largest film considered using Depletion+FS ($N_z = 24$), the predicted thermal conductivity matches the 3D-BTE-FS ($p=0$) result due to a Brillouin zone resolution that is the same as that used in the bulk calculation. As the film thickness is decreased, the thermal conductivity initially stays constant and then increases. It reaches a value close to the bulk thermal conductivity when $N_z=4$, which is clearly a non-physical result. The trend of increasing thermal conductivity with decreasing film thickness for the Depletion+FS calculation is consistent with the results of Turney et al., who studied LJ argon films at a temperature of 20 K.\cite{turney2010plane}
Although phonon depletion is an intuitive idea and a convenient way to consider phonon confinement, it remains an approximate 3D treatment that is not able to capture the correct trends or magnitudes of the thermal conductivity of ultrathin films.

The MD-GK, 2D-BTE, and 3D-BTE-FS thermal conductivity prediction methods were also applied at temperatures of 10 and 20 K. The results are plotted in Figs. S3 and \ref{F-thermalconductivity_argon}(b).
For MD-GK and 2D-BTE at both temperatures, the thermal conductivity plateau at very small thickness is again present, with the 2D-BTE results higher than those from MD-GK. 
Beyond a thickness of 10 unit cells at a temperature of 20 K, the MD-GK thermal conductivities unexpectedly become greater than the 2D-BTE values. This crossover, which we attribute to the use of the MD-derived structure in the 2D-BTE calculations, is discussed in Sec. S3.
To match the 2D-BTE predictions at a temperature of 10 K, the 3D-BTE-FS calculations require a specularity parameter of 0.34. At a temperature of 20 K, completely diffuse boundaries are sufficient to provide a match between the 2D-BTE and 3D-BTE-FS predictions.
As temperature increases, the atomic RMS displacements increase (Fig.~\ref{F-rms_all_argon}), making the surface rougher and decreasing the specularity parameter. The decrease of the fitted specularity parameter with increasing temperature follows this trend. Furthermore, the impact of the specularity parameter decreases with increasing temperature, where shorter intrinsic MFPs limit the reduction of the effective MFPs due to phonon-boundary scattering.

\subsection{\label{sec:argonplateau}Signatures of Confinement}

\begin{figure}[b]
\centering
\includegraphics{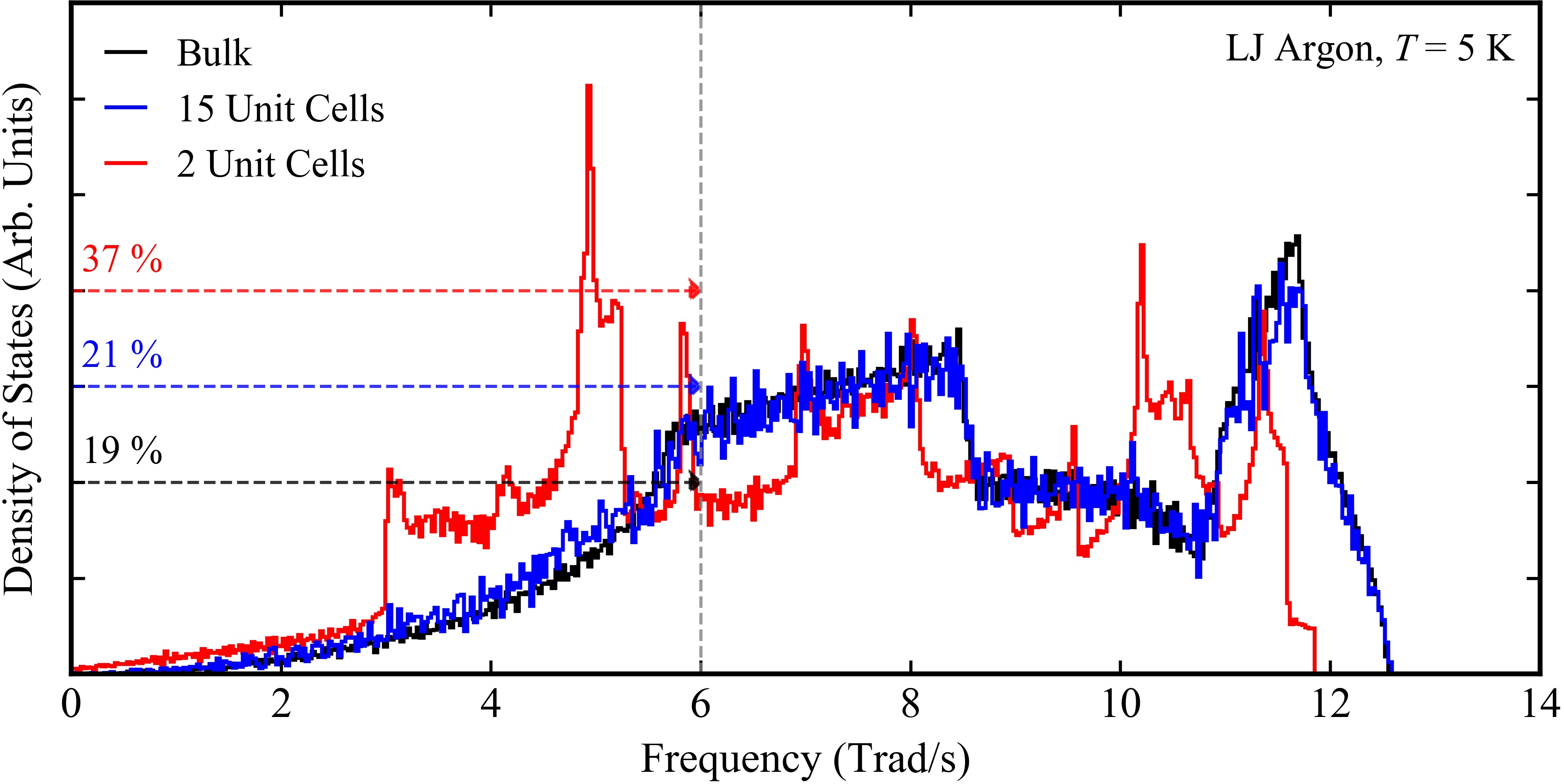}
\caption{Phonon DOS for the bulk, two unit cell film, and fifteen unit cell film argon systems at a temperature of 5 K. The vertical dashed line represents the end of the quadratic portion of the bulk DOS. Phonon modes with frequencies below this point are labeled as low-frequency. The percentages provided above the dashed horizontal lines indicate the contribution of the low-frequency phonons modes for that system. The DOS are normalized to have the same integrated area.}
\label{F-DOS_argon}
\end{figure}

We now analyze the phonon mode properties to identify signatures of confinement, with a focus on the 5 K system. 
The [100] phonon dispersion curves for the bulk and two unit cell film systems are plotted in Fig. S4.
The phonon DOS for the bulk, two unit cell film, and fifteen unit cell film systems are plotted in Fig.~\ref{F-DOS_argon}. The phonon dispersion and DOS were obtained by applying harmonic lattice dynamics calculations to the the time-averaged MD structures described in Sec.~\ref{sec:argonstr}. The film dispersions are characterized by the emergence of a quadratic acoustic flexural mode, as found in 2D materials. 
The bulk and fifteen unit cell film DOS are indistinguishable, with a maximum frequency of 12.5 Trad/s. The similarities in these two DOS suggests that the thermal conductivity reduction for the fifteen unit cell film is due to changes in the phonon scattering.
For the two unit cell film, however, the maximum frequency shifts downwards to 11.8 Trad/s and there is an increase in the DOS between frequencies of 3.0 and 5.2 Trad/s. 
Kim et al.~link these changes to the film's free surfaces, which remove the degeneracy of the transverse dispersion branches along high-symmetry directions.\cite{Kim2018mapping} 
The peak in the DOS at a frequency of 4.9 Trad/s in the two unit cell film corresponds to a flattened phonon branch, as shown in Fig. S4.

We label phonons modes with frequencies below 6 Trad/s as low-frequency. This cutoff is chosen because it corresponds to the end of the quadratic portion of the bulk DOS.
Low-frequency phonon modes in the two unit cell film account for 37\% of the total population. As the thickness increases, the percentage of low-frequency phonon modes decreases and approaches the bulk value of 19\%. 
We hypothesize that the large DOS at low frequencies in the very thin films is a signature of confinement.

\begin{figure}[tb]
\centering
\includegraphics{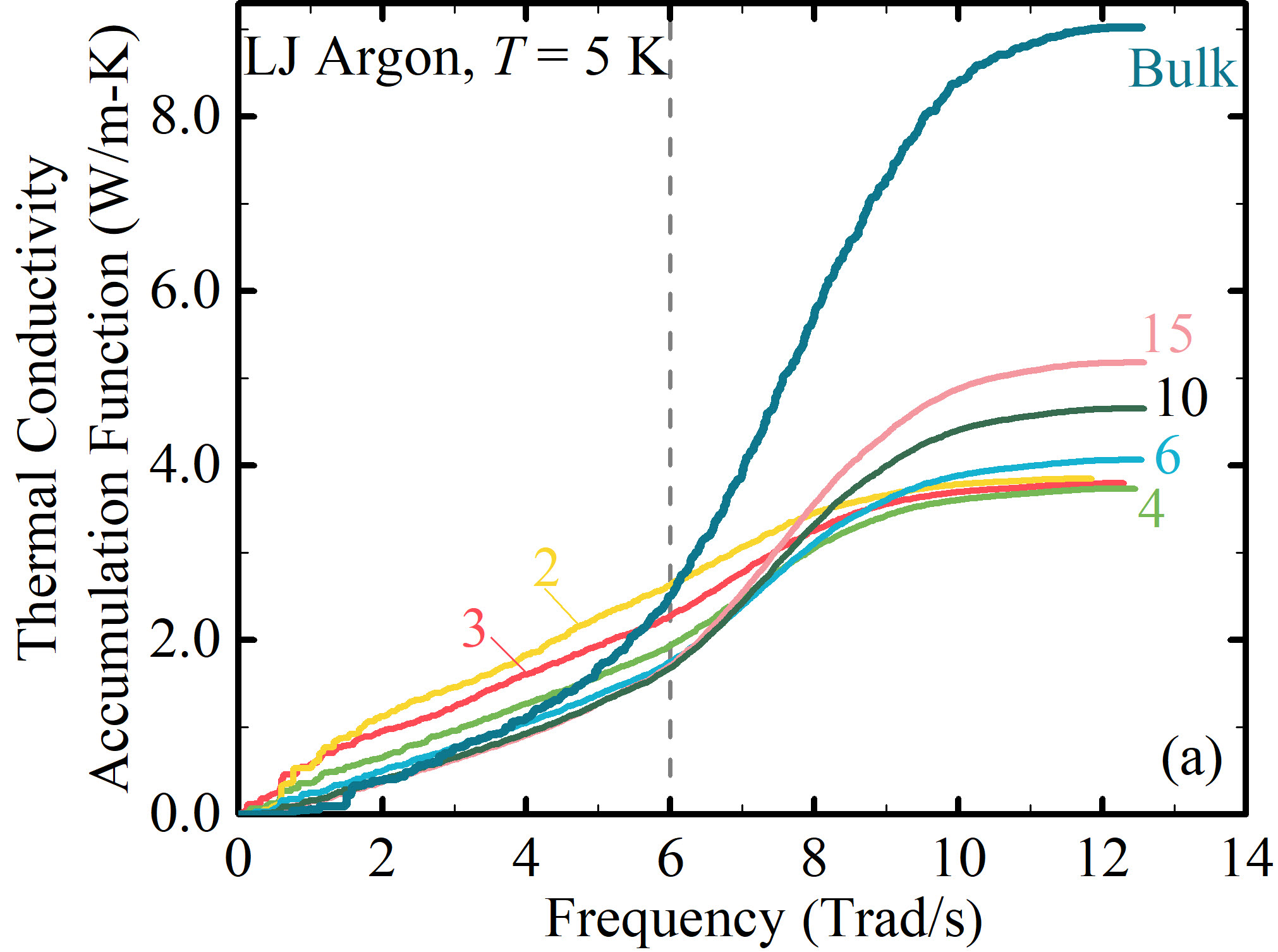}\\
\vspace{.2in}
\includegraphics{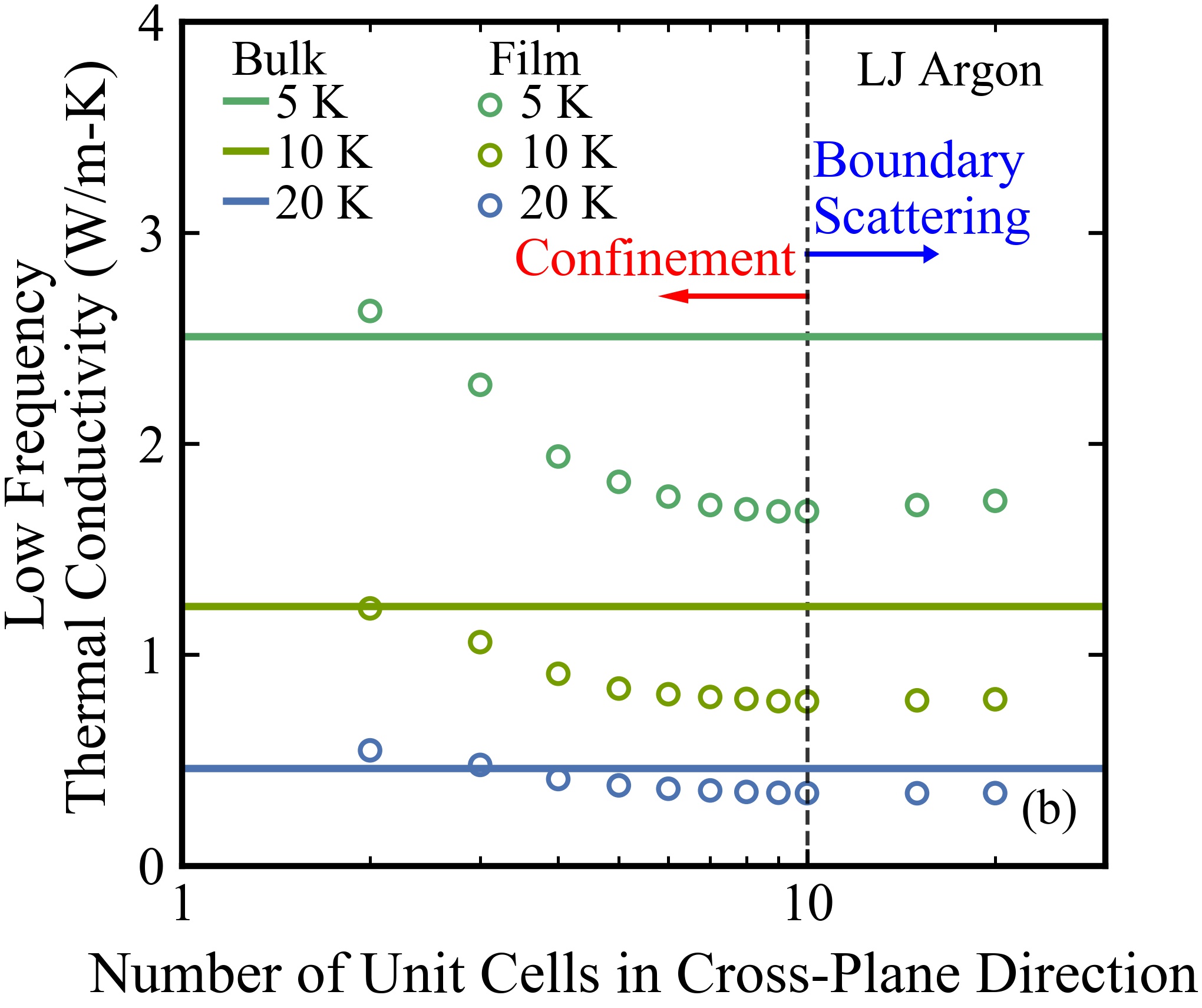}
\caption{(a) Argon frequency-dependent thermal conductivity accumulation functions for bulk and films predicted from 3D-BTE and 2D-BTE at a temperature of 5 K. The vertical dashed line at 6 Trad/s indicates the cutoff for low-frequency phonon modes. The numbers near the curves represents the film thickness in unit cells. (b) Accumulated thermal conductivity for phonon modes with frequencies below 6 Trad/s plotted versus film thickness. The horizontal solid line represents the bulk value. The vertical dashed line indicates the point where the accumulated thermal conductivity starts to increase with decreasing film thickness, which is independent of temperature.}
\label{F-accumulation_argon_freq}
\end{figure}

To test this hypothesis, the thermal conductivity accumulations as a function of frequency for bulk (3D-BTE) and for films (2D-BTE) with thicknesses between two and fifteen unit cells are plotted in Fig.~\ref{F-accumulation_argon_freq}(a).
The vertical coordinate of any point on the accumulation function represents the thermal conductivity contribution from phonon modes with frequencies less than the horizontal coordinate of that point (i.e., it is a cumulative distribution function). The vertical dashed line denotes the cutoff for the low-frequency phonon modes.

From bulk to films, the accumulation function at all frequencies decreases as the thickness decreases, particularly the contribution from high-frequency phonons, until the thickness reaches 10 unit cells. This trend is what would be expected based on a boundary scattering model like Fuchs-Sondheimer. Below this point, the accumulation functions in the low-frequency region increase with decreasing thickness. For the two-, three, and four unit cell films, the shape of the accumulation function is significantly different from that for bulk. 

To further analyze this behavior, the contribution of the low-frequency phonon modes to thermal conductivity is plotted in Fig.~\ref{F-accumulation_argon_freq}(b) as a function of film thickness. For bulk, phonon modes with frequencies below 6 Trad/s (i.e., what we have identified as low-frequency) contribute 2.5 W/m-K to the total thermal conductivity of 9.0 W/m-K, which is plotted as a solid horizontal line.
For the fifteen unit cell film, the contribution is 1.7 W/m-K, which is lower than bulk due to phonon-boundary scattering.
Below a thickness of ten unit cells, however, the thermal conductivity contribution from the low-frequency phonons increases as the film thickness decreases. The contribution reaches a value of 2.6 W/m-K for the two unit cell film, which is larger than that for bulk. This behavior cannot be predicted from a model that includes phonon-boundary scattering and is a clear indication of confinement. The increased contribution of low-frequency phonon modes directly correlates with the increased DOS of these modes, as shown in Fig.~\ref{F-DOS_argon}. Interestingly, as also shown in Fig.~\ref{F-accumulation_argon_freq}(b), the contribution of low-frequency phonon modes to the thermal conductivities of argon films at temperatures of 10 and 20 K also starts to increase below a thickness of ten unit cells. 
Recall that all of our lattice dynamics-based predictions are made using classical statistics so that they can be compared to the predictions from the MD simulations. If quantum statistics are used, such that low-frequency modes are more populated than high-frequency modes, the effect of confinement is even stronger, as shown in Fig. S5.

We conclude that, for argon, independent of temperature: (i) When the film thickness is less than ten unit cells, phonon confinement leads to an increased DOS and accumulated thermal conductivity for low-frequency modes. When this increase is sufficiently large, the total thermal conductivity plateaus. (ii) When the thickness is greater than ten unit cells, the confinement effect is limited and phonon-boundary scattering is dominant.

\clearpage

\section{\label{sec:silicon}Silicon Results}
\subsection{\label{sec:sistr}Film Atomic Structure}

\begin{figure}[b]
\centering
\includegraphics{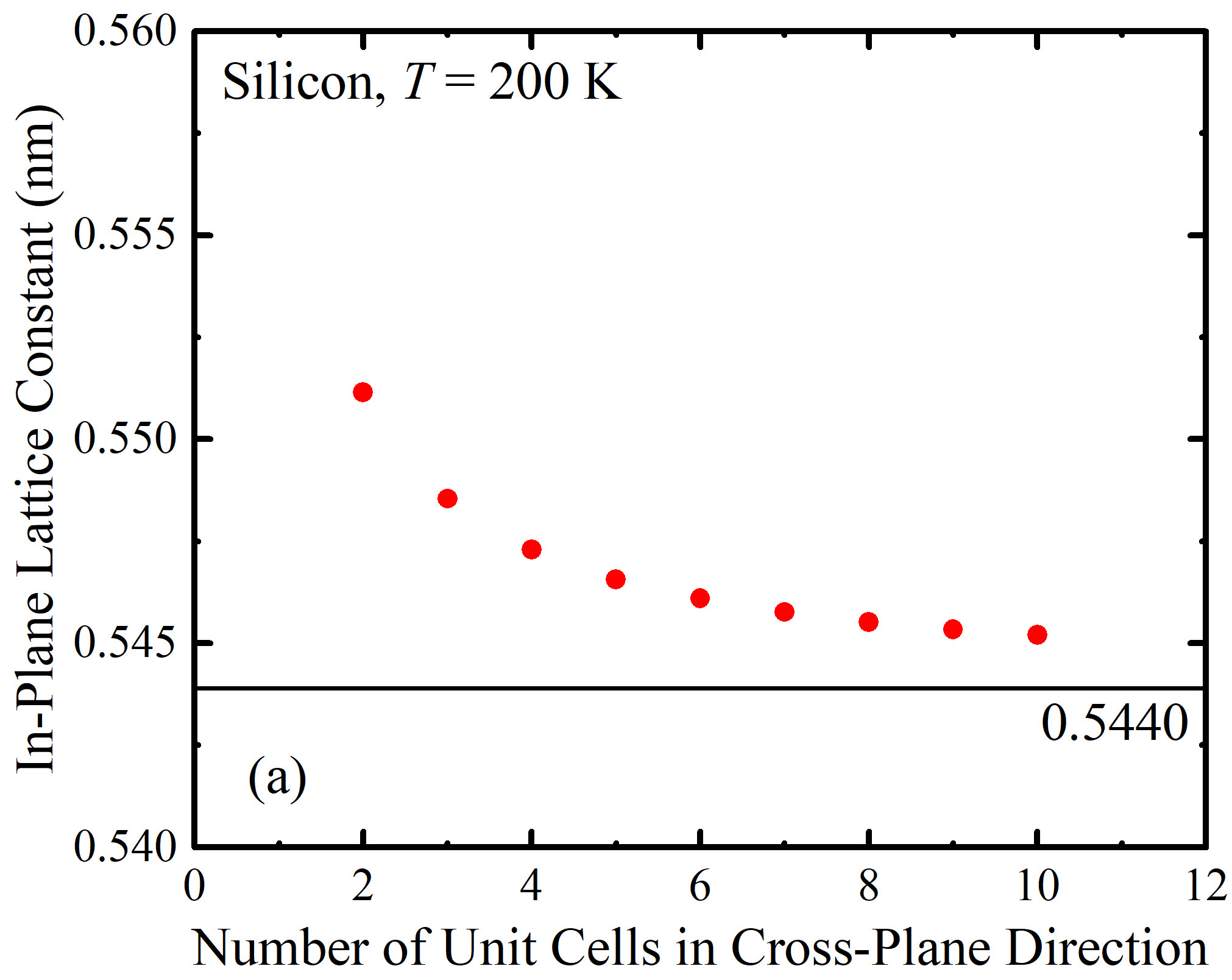}
\vspace{.2in}
\includegraphics{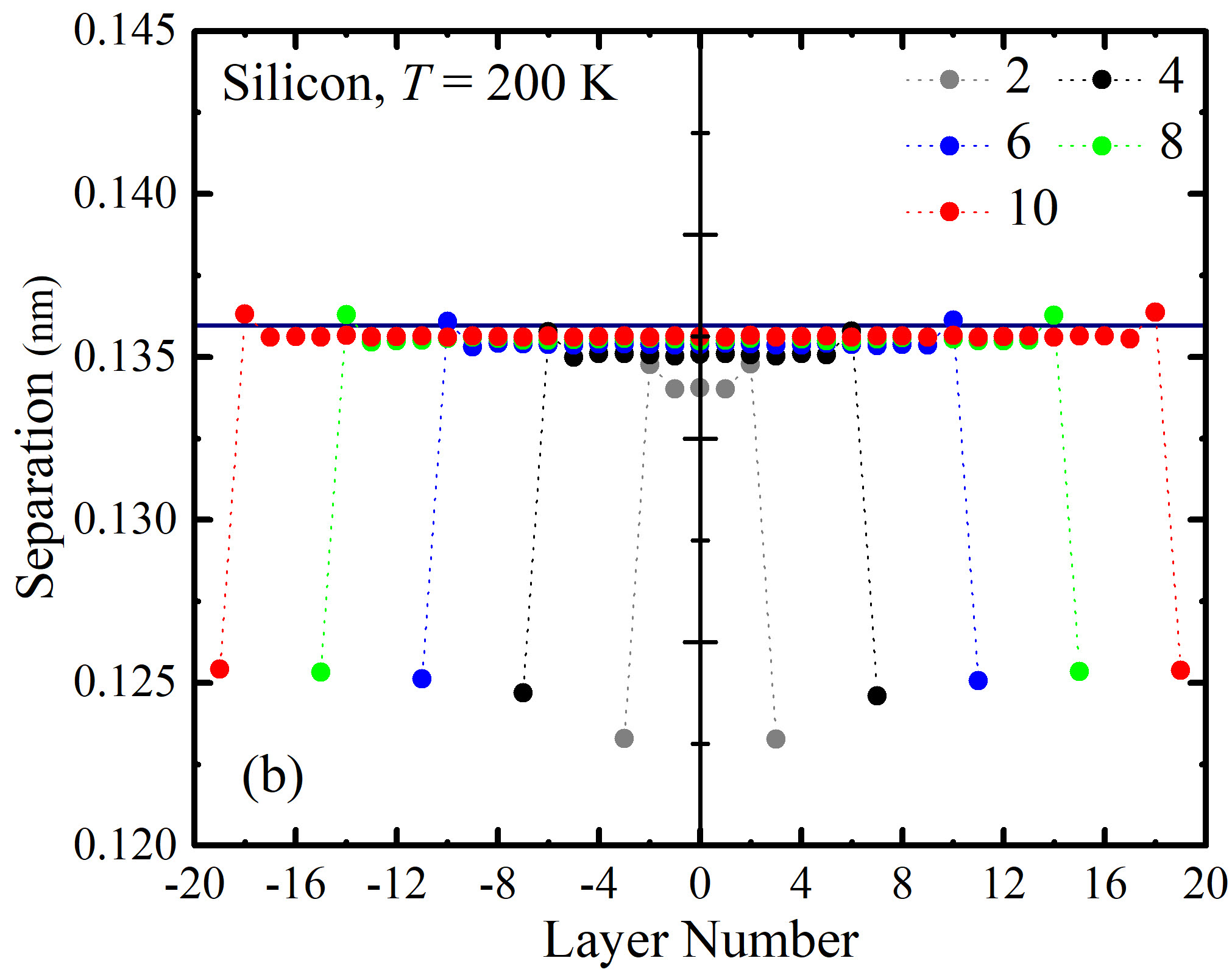}

\caption{(a) In-plane lattice constant versus film thickness for silicon films at a temperature of 200 K. The solid horizontal line represents the bulk value. (b) Separations between atomic layers in the cross-plane direction versus layer number. For the two unit cell film, there are eight layers and seven layer separations. ``0" corresponds to the separation between the two central layers while ``-3" and ``3" correspond to the separations involving the surface layers. The solid horizontal line represents the bulk value. The numbers in the legend correspond to the film thickness in conventional unit cells.}
\label{F-structure_si}
\end{figure}

The in-plane lattice constant for the silicon films is plotted in Fig.~\ref{F-structure_si}(a) for thicknesses between two and ten unit cells (1.0 - 5.5 nm). The bulk lattice constant (0.5440 nm) is plotted as a horizontal line. The films expand in the in-plane direction as the thickness is reduced, which is opposite to the argon result shown in Fig.~\ref{F-structure_argon}(a). The cross-plane layer separations are plotted in Fig.~\ref{F-structure_si}(b). The bulk layer separation is plotted as a horizontal line. 
For thicker films, similar to the argon results shown in Fig.~\ref{F-structure_argon}(b), the layers around the centerline are bulk-like.
As the thickness decreases, the deviation from bulk at the centerline increases to 1.5\% for the two unit cell film, larger than that of 1\% observed in Fig.~\ref{F-structure_argon}(b) for argon. 
At the surfaces, the layer separation is smaller than the bulk value, which is opposite to the findings for argon. Only the outermost layer shows a contraction, while the second outermost layer has a separation that is 0.2\% larger than bulk. The contraction at the surface is a result of strong attractive forces from the first few neighbor shells, which originate from the stiffness of the Tersoff potential.
The silicon and argon films are thus both anisotropic, but with changes in opposite directions compared to the bulk. 

\subsection{\label{sec:sianh}Anharmonicity}

\begin{figure}[b]
\centering
\includegraphics{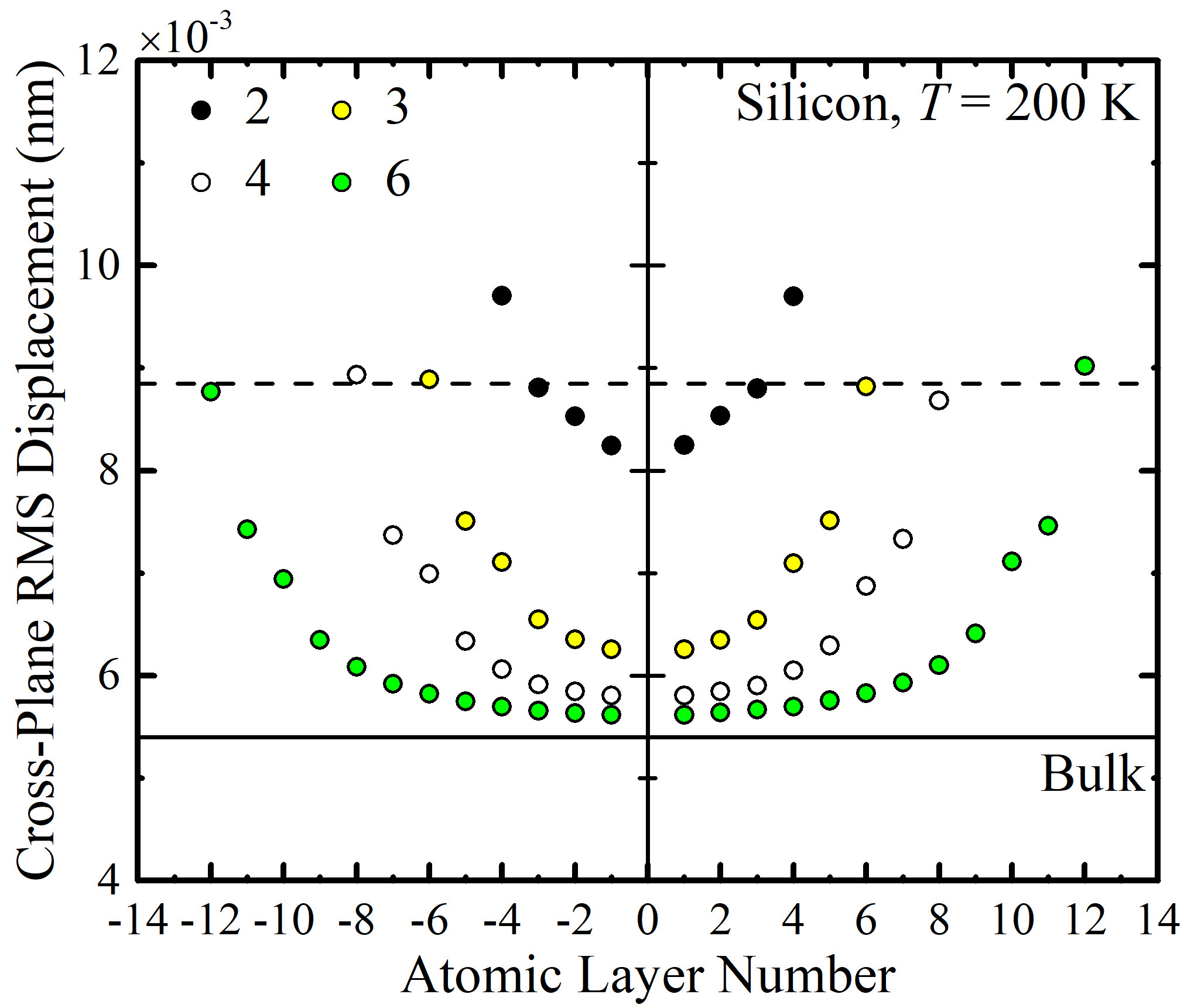}
\caption{Cross-plane RMS displacement versus atomic layer number for silicon films at a temperature of 200 K. The solid line represents the bulk RMS displacement along one Cartesian direction. The dashed line represents the value that the surface atoms approach in the cross-plane direction. The numbers in the legend correspond to the film thickness in conventional unit cells.}
\label{F-rms_cross}
\end{figure}

The cross-plane atomic RMS displacements for silicon films with thicknesses between two and six unit cells are plotted versus layer number in Fig.~\ref{F-rms_cross} along with the bulk value. The in-plane values are provided in Fig. S6. Similar to argon, both surface atoms and interior atoms in films with small thickness have larger RMS displacements than in bulk, which is an indication of stronger anharmonicity. The RMS displacements of the surface atoms are larger than those of the interior atoms, even though their layer separation is smaller [Fig.~\ref{F-structure_si}(b)]. 
As the thickness increases, the RMS displacement of the atoms in the central interior layer approach the bulk value of 0.0054 nm, while that for the surface atoms reach 0.0088 nm~(shown as a dashed horizontal line), which is 67\% larger. 
The expansion of the surface layers in argon and their contraction in silicon both increase anharmonicity. This observation is consistent with the findings of Parrish et al., who found an increase in anharmonicity induced by tension in bulk argon and by compression in bulk silicon.\cite{parrish2014origins} 

\subsection{\label{sec:sik}Thermal Conductivity}
\begin{figure}[b]
\centering
\includegraphics{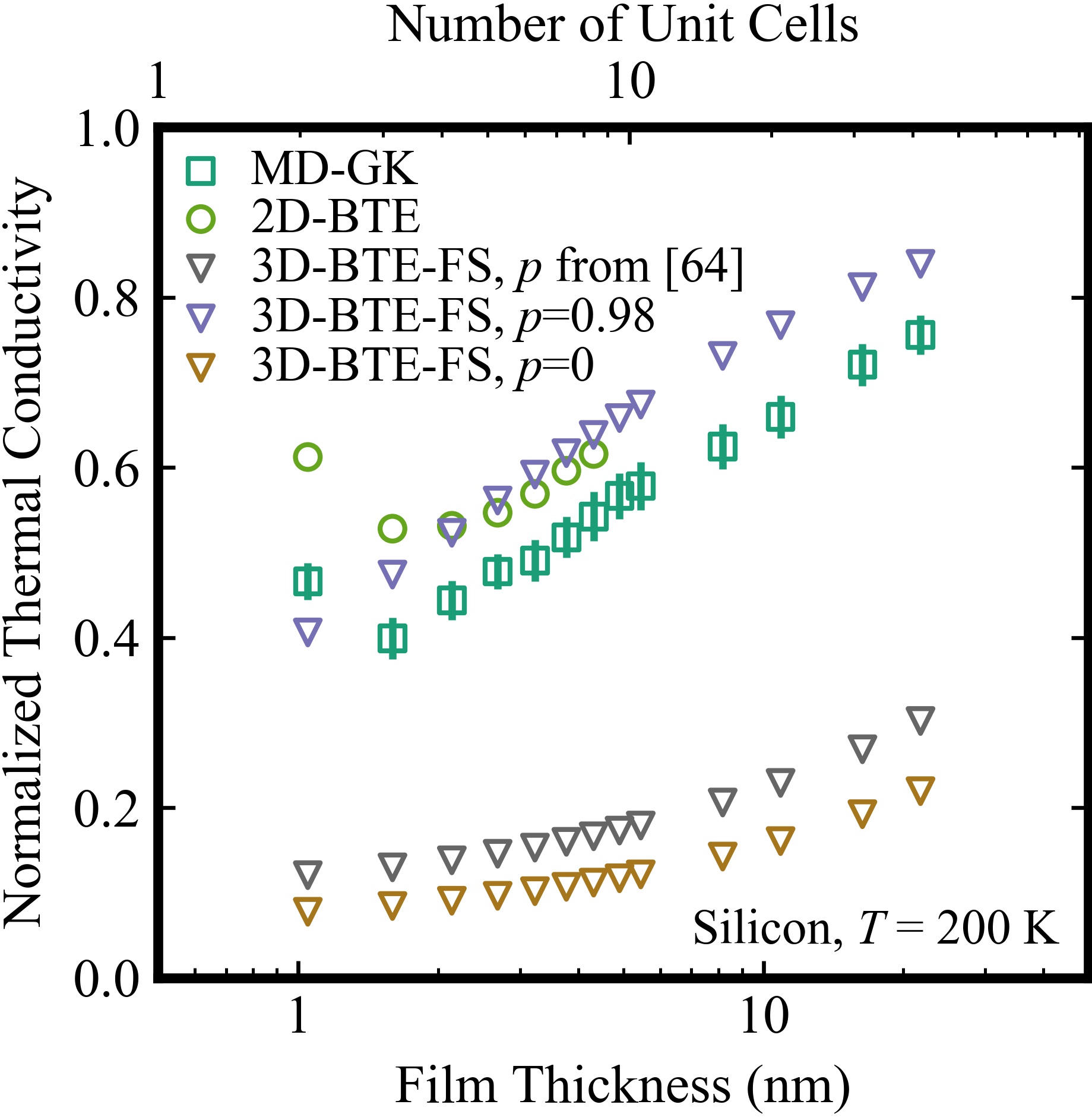} 
\caption{Thermal conductivity versus thickness of silicon films at a temperature of 200 K. 
The results from the MD-GK, 2D-BTE, and 3D-BTE (with bulk phonons combined with the Fuchs-Sondheimer boundary scattering model) methods are presented. The film thermal conductivities from the 2D-BTE and 3D-BTE methods are normalized by the bulk value from 3D-BTE (464 W/m-K), while that from MD-GK is normalized by the bulk MD-GK value (414 W/m-K). The Fuchs-Sondheimer boundary scattering model is applied with constant ($p =$ 0 and 0.98) and wavelength-dependent\cite{Ravichandran:2018eff} specularity parameters. }
\label{F-thermalconductivity_si}
\end{figure}

The thermal conductivities of the silicon films predicted from the MD-GK, 2D-BTE, and 3D-BTE-FS methods are plotted in Fig.~\ref{F-thermalconductivity_si}. 
To allow for a better comparison with Fig.~\ref{F-thermalconductivitydata}, the thermal conductivities are scaled by their corresponding bulk values: 414 W/m-K for MD-GK and 464 W/m-K for 3D-BTE. The bulk predictions are higher than the experimental value of 293 W/m-K.\cite{inyushkin2004isotope} 
The Tersoff potential is well-known for overpredicting thermal conductivity.\cite{broido2005lattice} The raw thermal conductivities are plotted in Fig. S7.

As the thickness decreases, both MD-GK and 2D-BTE predict a decreasing thermal conductivity. There is a minimum at a thickness of three unit cells, below which the thermal conductivity shows an unexpected increase of about 15\%. 
This behavior is similar to the argon data plotted in Fig.~\ref{F-thermalconductivity_argon}, but with an increase instead of a plateau. This result suggests that the phonon confinement effect is stronger in silicon, a hypothesis that we further explore in Sec.~\ref{sec:sisig} by considering the mode-dependent phonon properties. 

As expected, the 2D-BTE thermal conductivities are always higher than the MD-GK values. 
The deviation increases as the film thickness decreases and reaches a maximum of 91 W/m-K for the 2 unit cell film, as shown in Fig. S7.
The increase in the deviation is analogous to that in argon as discussed in Sec.~\ref{sec:argonk}, and is consistent with the increase of the anharmonicity as the film gets thinner, as shown in Fig.~\ref{F-rms_cross}.

The scaled thermal conductivities predicted from 3D-BTE-FS ($p=0$) (i.e., with a specularity parameter of zero) vary from 0.08 (two unit cell film) to 0.22 (forty unit cell film), which is much lower than 2D-BTE scaled values, which range from 0.53 to 0.62 over the computationally accessible range of two to eight unit cells. Based on the argon results, the potential role of specular phonon-boundary interactions must be assessed. We first consider the experimentally-derived wavelength-dependent specularity parameters for silicon reported by Ravichandran et al.~at a temperature of 300 K.\cite{Ravichandran:2018eff} We use the upper bound of their results, which vary from 0.18 to 1 as the wavelength increases.
Applying their results to the 3D-BTE-FS model increases thermal conductivity, leading to scaled values ranging from 0.12 (two unit cell film) to 0.30 (forty unit cell film). These values remain smaller than the exact 2D-BTE treatment. To achieve a good agreement between 3D-BTE-FS and 2D-BTE, a mode-independent specularity parameter of 0.98 is required, which is non-physical at finite temperatures for the full phonon spectrum. 
The thermal conductivity increase at the smallest thickness is not observed with any choice of specularity parameter, consistent with the results for argon. 

The scaled thermal conductivities predicted from MD-GK range from 0.36 [the minimum value at three unit cells (1.6 nm)] to 0.76 [for the maximum thickness considered of forty unit cells (21.8 nm)]. 
Neogi et al.\cite{neogi2015tuning} applied MD-GK to Tersoff silicon films with reconstructed $2 \times 1$ surfaces. They predicted scaled thermal conductivities at a temperature of 300 K that varied from 0.37 (1 nm) to 0.71 (20 nm), consistent with our results. 
Their MD-GK predictions for these silicon films were ten times larger than their experimental measurements (green triangles in Fig.~\ref{F-thermalconductivitydata}) and those of Ch\'avez-\'Angel et al. (black triangles in Fig.~\ref{F-thermalconductivitydata}). \cite{chavez2014reduction}  
By including surface roughness and a native oxide layer in their MD simulations, they obtained scaled thermal conductivities that agreed with the experimental measurements. They then etched the experimental oxide layer and measured increased thermal conductivities (blue triangles in Fig.~\ref{F-thermalconductivitydata}), which agreed with their MD simulations of rough films without the oxide. 
As reported by Xiong et al.,\cite{Xiong:20176b2} the native oxide layer can introduce local resonant modes that interact with the propagating phonons inside the film, reducing their group velocity and lowering thermal conductivity. This phenomenon is analogous to predictions for nanophononic metamaterials.\cite{honarvar2018two,Davis:2014d32}

As shown in Fig.~\ref{F-thermalconductivitydata}, thermal conductivity predictions from 3D-BTE-FS ($p=0$) (corresponding to the FS+Esfarjani curve) agree with the experimental TTG data for suspended silicon films with thicknesses as small as 20 nm. 3D-BTE-FS does not include the effects of confinement, however, and, as we showed in Fig.~\ref{F-thermalconductivity_si}, leads to a large underestimation of thermal conductivity compared to the rigorous 2D-BTE and MD-GK approaches, even with realistic specularity parameters. 
As found by Neogi et al.,\cite{neogi2015tuning} the low thermal conductivities of suspended silicon films measured by Raman thermometry with thicknesses smaller than $\sim 200$ nm (green and black triangles in Fig.~\ref{F-thermalconductivitydata}) result from the effects of confinement and the presence of a native oxide layer. None of the Raman thermometry measurements are well predicted by 3D-BTE-FS, which does not capture thermal transport inside the oxide and/or the impact of the oxide on the film's phonon properties. The same holds true for the suspended island measurements (red and pink squares). It is thus possible that the reasonable agreement found 
between 3D-BTE-FS ($p=0$) and the TTG experimental measurements for ultrathin silicon films is coincidental due to use of bulk phonon properties with a boundary scattering modeling and ignoring the oxide layer.

\subsection{\label{sec:sisig} Signatures of Confinement}

\begin{figure}[b]
\centering
\includegraphics{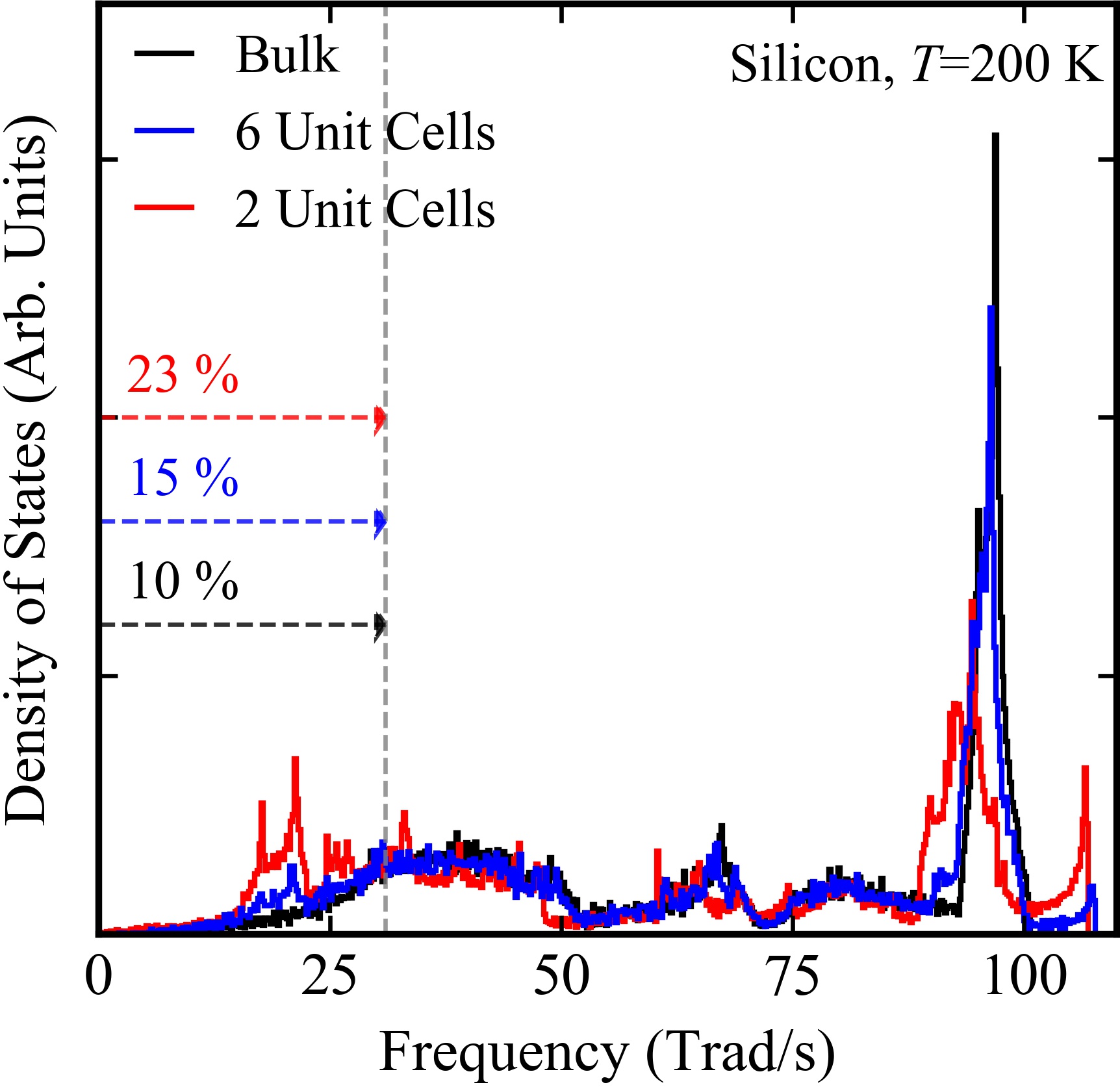}
\caption{Silicon DOS for bulk, the two unit cell film, and the six unit cell film using the atomic structures at a temperature of 200 K. The vertical dashed line indicates the end of the quadratic portion of the bulk DOS. Phonon modes with frequencies below this point are labeled as low-frequency. The percentages provided above the horizontal dashed lines indicate the contribution of low-frequency phonon modes to the DOS for that system.}
\label{F-DOS_si}
\end{figure}

Following the argon analysis presented in Sec.~\ref{sec:argonplateau}, the DOS of bulk and silicon film systems (two and six unit cells) are plotted in Fig.~\ref{F-DOS_si}. The quadratic portion of bulk DOS ends at a frequency of 31 Trad/s, which we identify as the low-frequency region. There are two peaks between frequencies of 20 and 25 Trad/s in the films that are not present in bulk, similar to argon. For the two unit cell film, 23\% of the phonon modes are low-frequency, which is more than double that for bulk.

\begin{figure}[tb]
\centering
\includegraphics{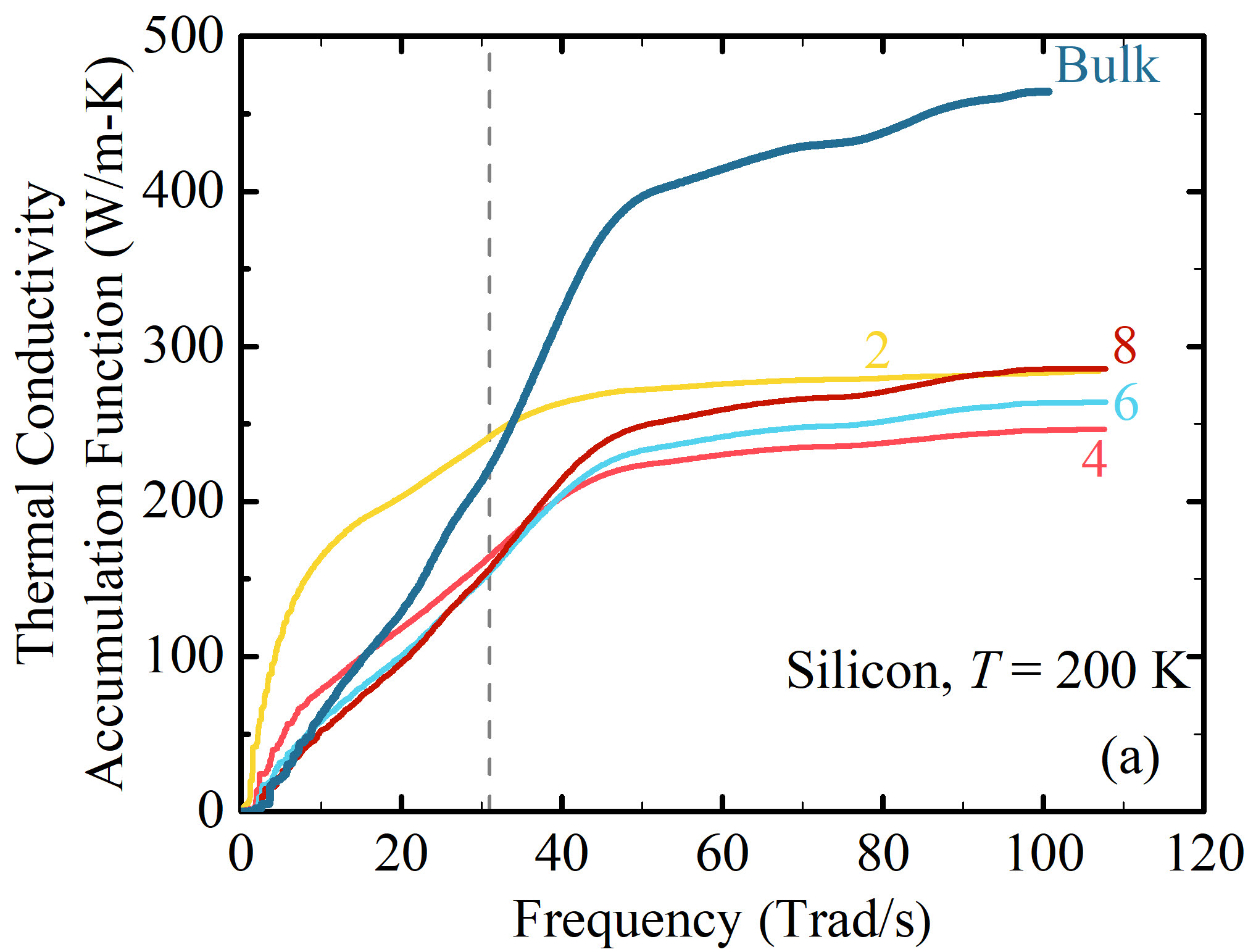}\\
\vspace{0.1in}
\includegraphics{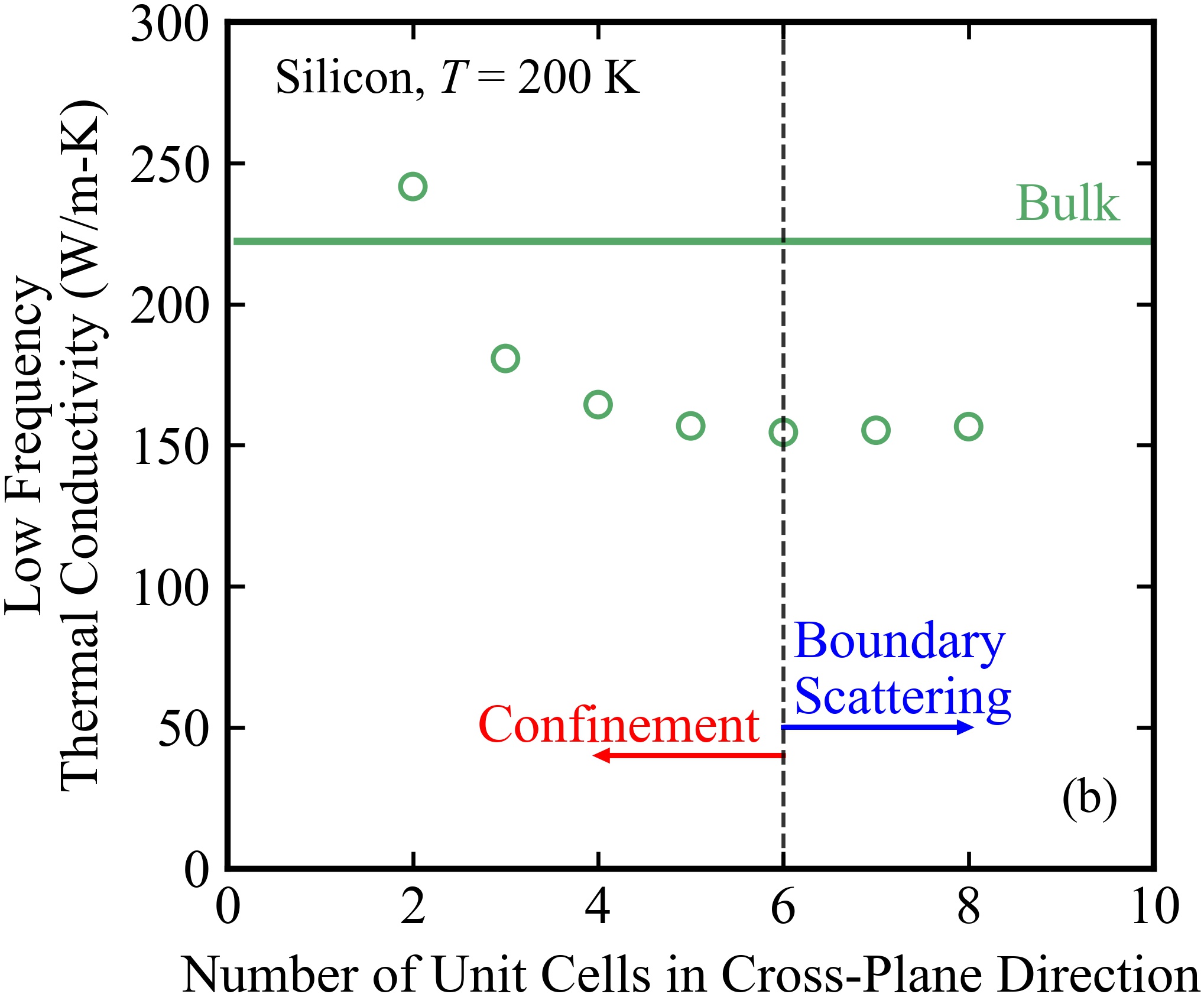}
\caption{(a) Silicon frequency-dependent thermal conductivity accumulation functions for bulk and films predicted from 3D-BTE and 2D-BTE at a temperature of 200 K. The vertical dashed line at 31 Trad/s indicates the cutoff for low-frequency phonon modes. (b) Accumulated thermal conductivity for phonon modes with frequencies below 31 Trad/s plotted versus film thickness. The horizontal solid line represents the bulk value. The vertical dashed indicates the point where the accumulated thermal conductivity starts to increase with decreasing film thickness.}
\label{F-accumulation_si}
\end{figure}

The frequency-dependent thermal conductivity accumulation functions for silicon bulk and films are plotted in Fig.~\ref{F-accumulation_si}(a). In both bulk and films, phonon modes with low frequencies make the dominant contribution to thermal transport. In moving from bulk to the films, the accumulation function decreases at all frequencies as the thickness decreases, until the thickness reaches six unit cells. Below this point, the contributions from phonon modes with frequencies below 31 Trad/s (i.e. low-frequency phonon modes) increase as the thickness is further decreased. This trend is similar to that observed in argon. Furthermore, the accumulation function for the two unit cell film is significantly different compared to bulk. 

The contribution of low-frequency phonon modes to thermal conductivity is plotted in Fig.~\ref{F-accumulation_si}(b).
For bulk, the low-frequency phonon modes contribute 222 W/m-K to the total thermal conductivity of 464 W/m-K. For the eight unit cell film, the contribution is 157 W/m-K, which is lower than bulk due to phonon-boundary scattering. Below a thickness of six unit cells, the contribution from the low-frequency phonon modes increases as the film thickness decreases and they exhibit a larger DOS. The contribution reaches a value of 242 W/m-K for the two unit cell film, which is larger than that for bulk. This trend is similar to that observed in argon [Fig.\ref{F-accumulation_argon_freq}(b)]. Interestingly, the transition point in silicon is 24 atomic layers, which is close to the 20 atomic layers (10 unit cells) observed for argon. 

Low-frequency phonon modes account for 48\% of the bulk silicon thermal conductivity and 28\% of the bulk argon thermal conductivity. For the two unit cell films, the contribution of the low-frequency phonon modes is 85\% for silicon and 67\% for argon. We believe that the greater contribution for the silicon films is what leads to the increase in their total thermal conductivity for the smallest thicknesses, compared to the plateau observed for argon.

\section{\label{sec:sum}Summary}
We investigated in-plane thermal transport by phonons in ultrathin films of (i) LJ argon at temperatures of 5, 10, and 20 K, and (ii) Tersoff silicon at a temperature of 200 K using MD simulations, lattice dynamics calculations based on 2D and 3D Brillouin zones, and the BTE. The finite-temperature atomic structures used for all thermal conductivity predictions were generated from MD simulations. As shown in Figs.~\ref{F-structure_argon} and \ref{F-structure_si}, there is anisotropy in the in-plane and cross-plane atomic spacings in the films compared to the corresponding isotropic bulk structures. RMS atomic displacement obtained from MD simulations revealed that surface atoms experience a stronger anharmonicity than those at the film center, as shown in Figs.~\ref{F-rms_all_argon} and \ref{F-rms_cross}, which has important consequences when comparing the MD-GK and  2D-BTE thermal conductivity predictions. 

For argon modeled with MD and 2D lattice dynamics, a thermal conductivity plateau is observed for films with thicknesses smaller than four unit cells (2.1 nm) (Fig.~\ref{F-thermalconductivity_argon}). For the same calculations for silicon, the film thermal conductivity increases when its thickness is reduced below three unit cells (1.6 nm) (Fig.~\ref{F-thermalconductivity_si}). We determined that these unexpected trends are a signature of phonon confinement, which causes an increase in the DOS of low-frequency phonons and a corresponding increase in their contribution to thermal conductivity to a level that can exceed that present in the bulk structures, as shown in Figs.~\ref{F-accumulation_argon_freq}(b) and \ref{F-accumulation_si}(b). This effect could potentially be observed experimentally by applying emerging techniques for measuring phonon mode-dependent contributions to thermal conductivity.\cite{regner2015advances}
 
Bulk phonon properties combined with the Fuchs-Sondheimer model for phonon-boundary scattering were also used to predict the film thermal conductivities (i.e., 3D-BTE-FS). These predictions did not show the plateau (for argon) or the increase (for silicon) in thermal conductivity at very small thicknesses. The common assumption of fully diffuse phonon-boundary scattering was not able to capture the thermal conductivity magnitudes for thicker argon films at temperatures of 5 and 10 K or for the silicon film at a temperature of 200 K. We also showed that the phonon depletion method produces non-physical thermal conductivities. 

Our work demonstrates that the rigorous 2D description of phonons in ultrathin films is essential for accurate thermal conductivity prediction and for elucidating the underlying transport physics. As advances continue to be made for the fabrication and characterization of ultrathin films, and notably suspended films,\cite{cuffe2015reconstructing, neogi2015tuning, chavez2014reduction, wang2017fundamental} the tools described herein will thus be of utmost importance.

\begin{acknowledgements}

 B.~F. and G.~T. acknowledge financial support from the National Natural Science Foundation of China under grant number 51825604. K.~D.~P., H.-Y.~K., and A.~J.~H.~M. acknowledge financial support from the National Science Foundation Award DMR-1507325. B.~F. acknowledges financial support from a program of the China Scholarships Council (Award No.~201706280251). G.~T. acknowledges financial support from the 111 Project under grant number B16038. 

\end{acknowledgements}

\bibliographystyle{apsrev4-1} 
\bibliography{Ref0303.bib} 

\end{document}


\preprint{}

\title{Supplemental Material for \\Phonon Confinement and Transport in Ultrathin Films}

\author{Bo Fu\begin{CJK*}{UTF8}{gbsn}
(傅博)
\end{CJK*}}
\affiliation{MOE Key Laboratory of Thermo-Fluid Science and Engineering, School of Energy and Power Engineering, Xi'an Jiaotong University, Xi'an 710049, China}
\affiliation{Department of Mechanical Engineering, Carnegie Mellon University, Pittsburgh, PA 15213, USA}
\author{Kevin D. Parrish}
\affiliation{Department of Mechanical Engineering, Carnegie Mellon University, Pittsburgh, PA 15213, USA}
\author{Hyun-Young Kim}
\affiliation{Department of Mechanical Engineering, Carnegie Mellon University, Pittsburgh, PA 15213, USA}
\author{Guihua Tang\begin{CJK*}{UTF8}{gbsn} (唐桂华) \end{CJK*}} \email{ghtang@mail.xjtu.edu.cn.}
\affiliation{MOE Key Laboratory of Thermo-Fluid Science and Engineering, School of Energy and Power Engineering, Xi'an Jiaotong University, Xi'an 710049, China}
\author{Alan J. H. McGaughey} \email{mcgaughey@cmu.edu.}
\affiliation{Department of Mechanical Engineering, Carnegie Mellon University, Pittsburgh, PA 15213, USA}

\date{\today}

\maketitle

\section{Lennard-Jones Argon Results}
\begin{figure}[h]
\centering
\includegraphics{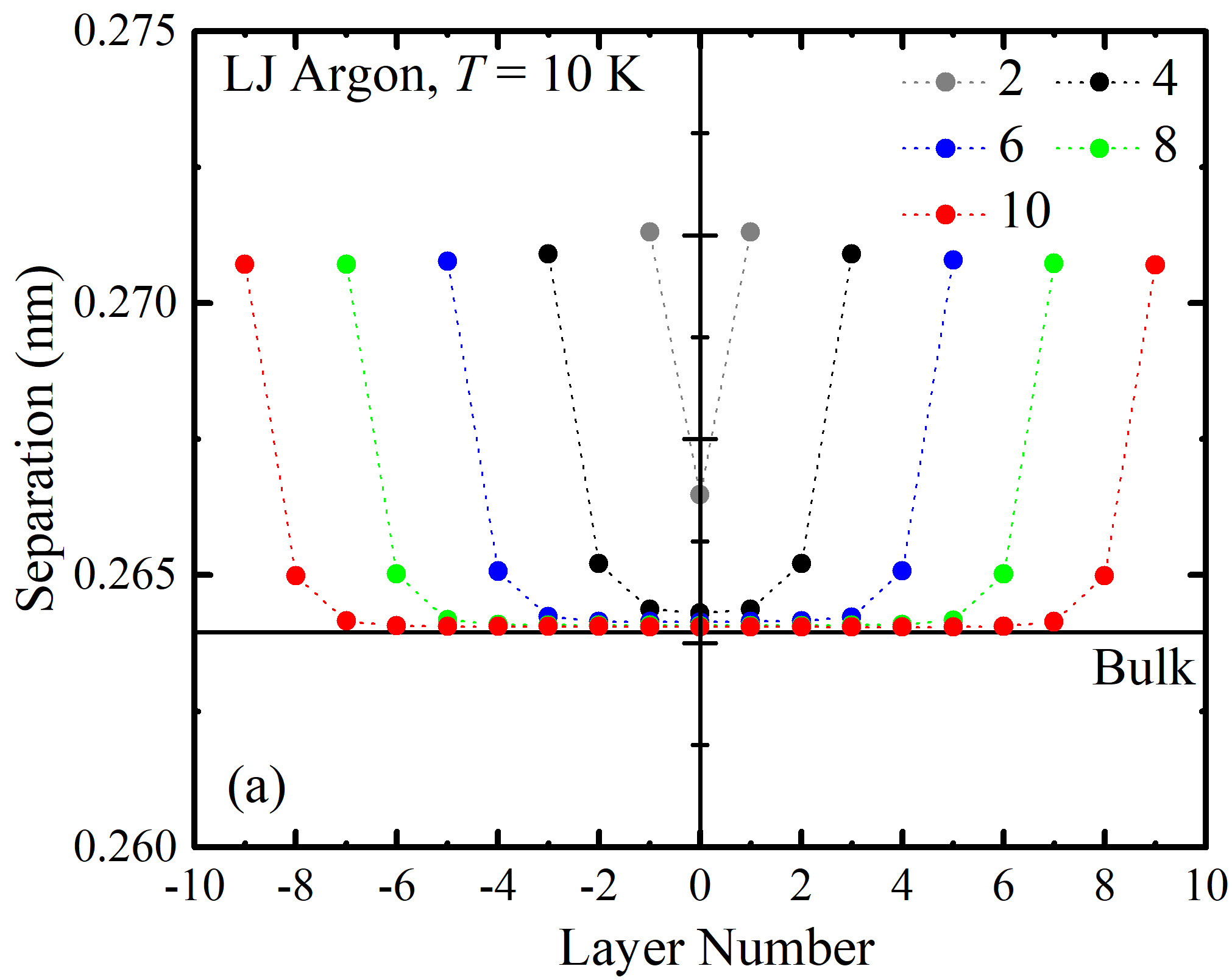}\\
\vspace{.1in} 
\includegraphics{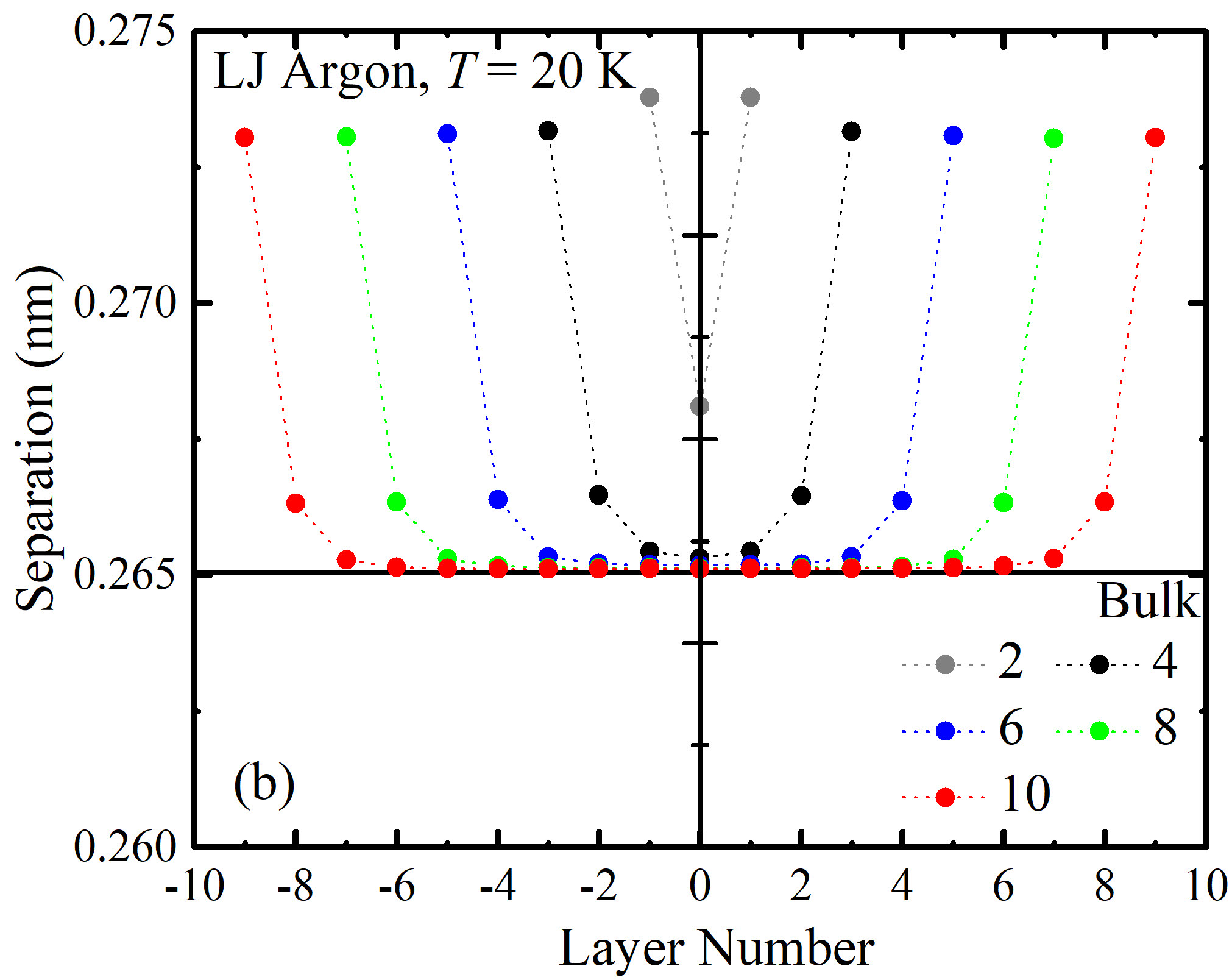}
\caption{Separations between atomic layers for argon films in the cross-plane direction at temperatures of (a) 10 K and (b) 20 K versus the layer number, which represents the location in the film. 
For example, for the two unit cell film, there are four layers and three layer separations. ``0" corresponds to the separation between the center layers while ``-1" and ``1" correspond to the separations involving the surface layers. The solid horizontal line represents the bulk values. The numbers in the legend correspond to the film thickness in conventional unit cells.}
\label{F-structure_argon_1020K}
\end{figure}
\newpage

\begin{figure}[h]
\centering
\includegraphics{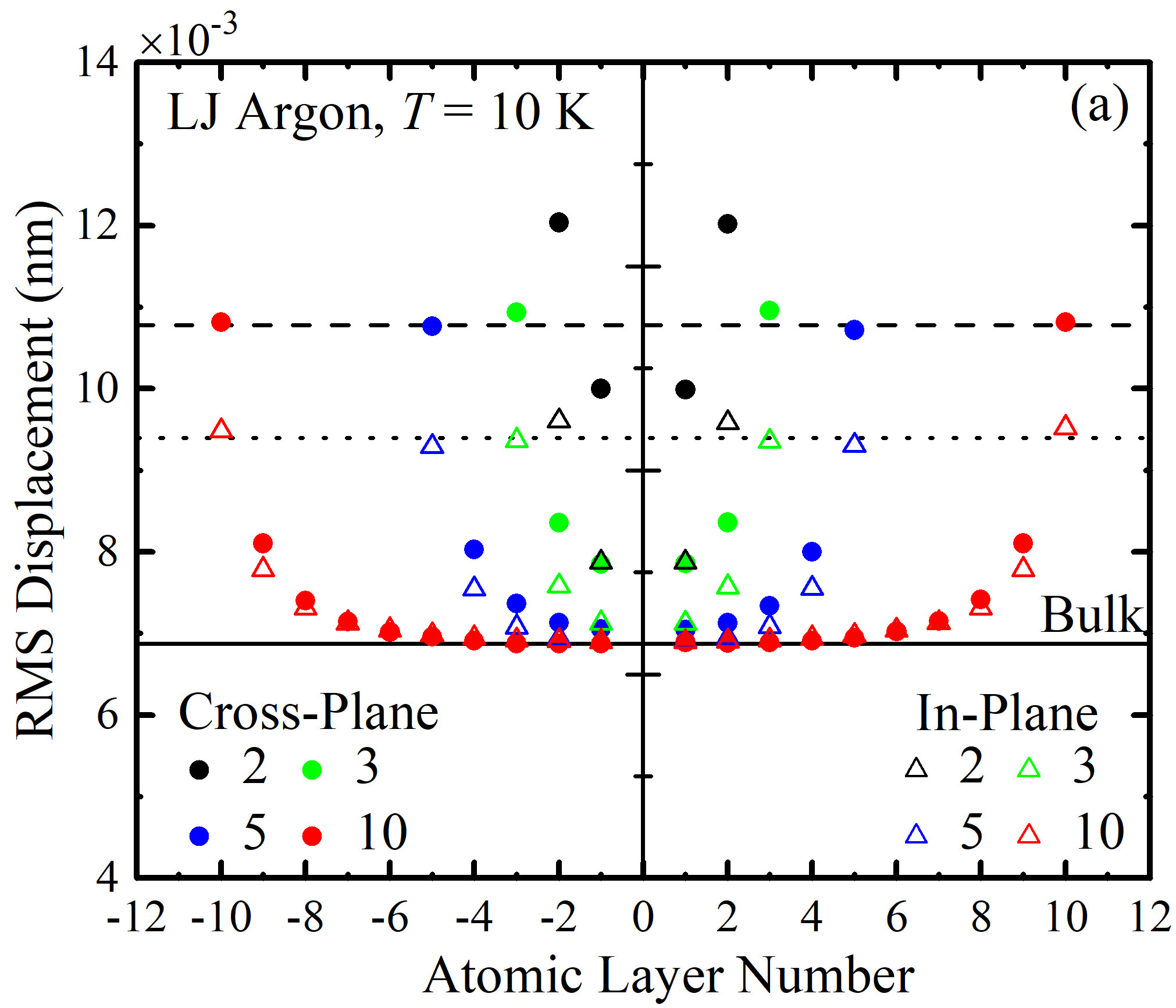}\\
\vspace{.1in} 
\includegraphics{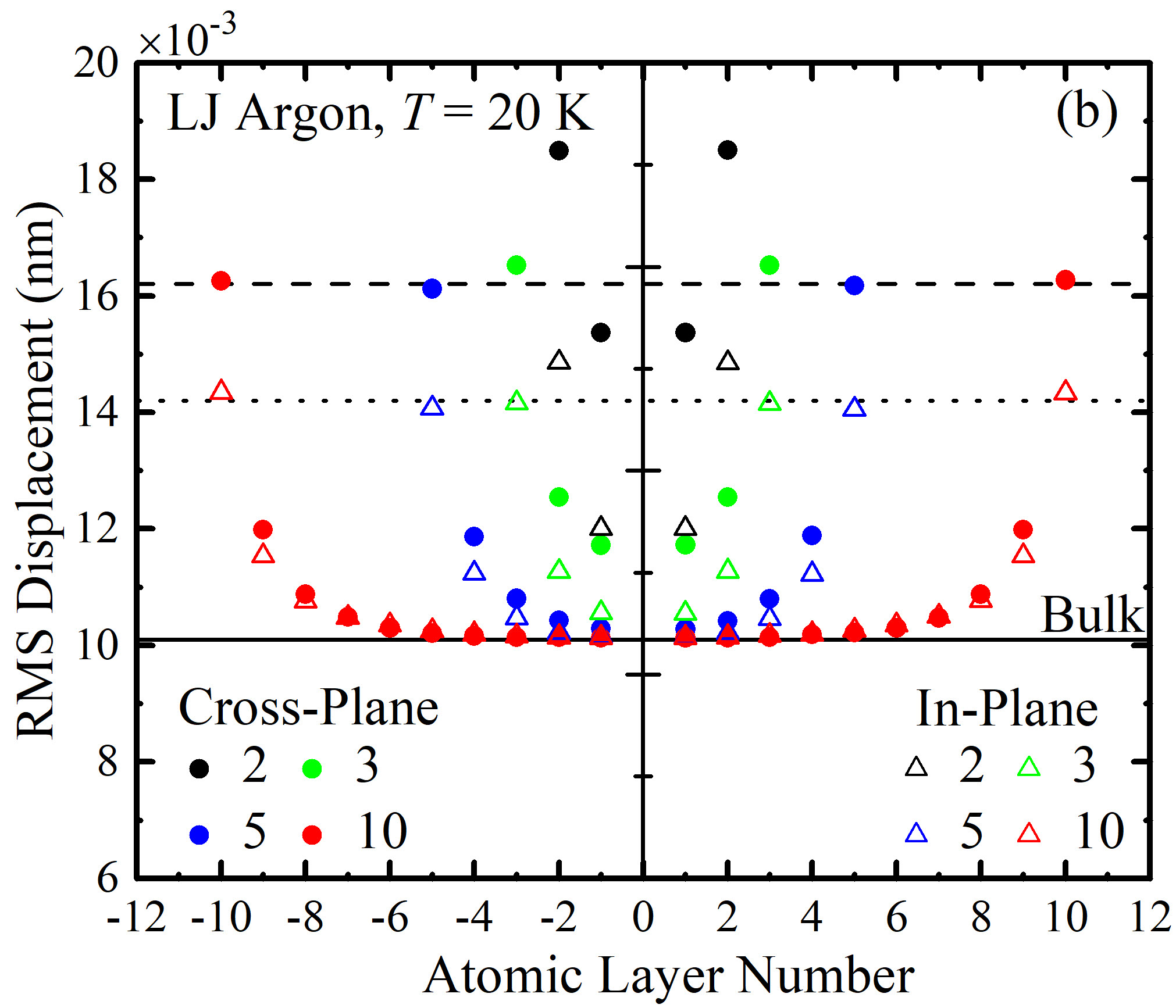}
\caption{RMS displacement in the cross-plane and in-plane directions versus atomic layer number for argon films at temperatures of (a) 10 K and (b) 20 K. The solid line represents the bulk RMS displacement along one Cartesian direction. The in-plane RMS displacements are averaged over the $x$- and $y$-directions. The dashed and dotted lines represent the values that the surface atoms approach in the cross-plane and in-plane directions. The numbers in the legend correspond to the film thickness in conventional unit cells.}
\label{F-rms_all_argon_1020K}
\end{figure}
\newpage

\begin{figure}[h]
\centering
\includegraphics{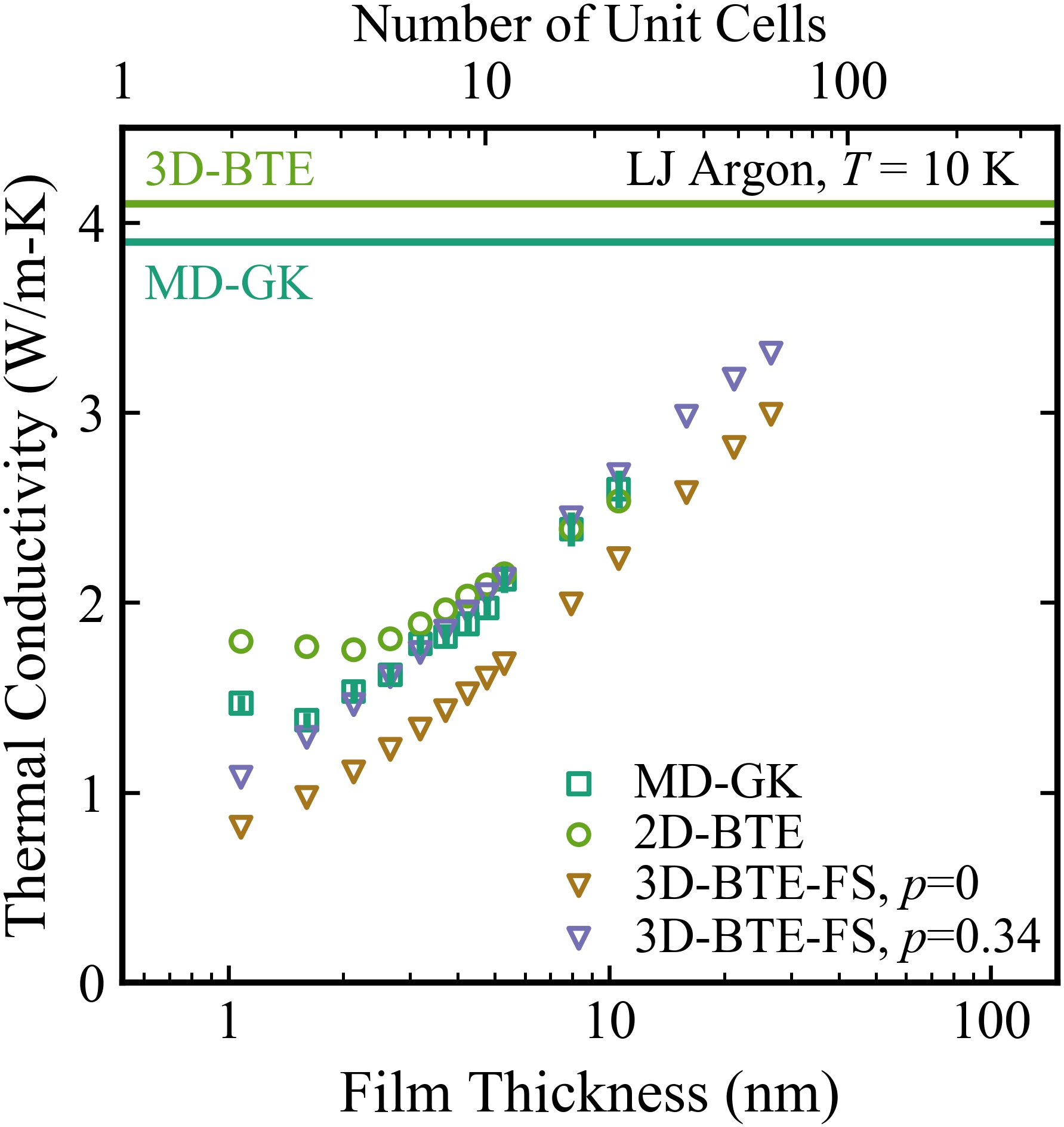}
\caption{Thermal conductivity versus thickness of argon films at a temperature of 10 K. 
The results from the MD-GK, 2D-BTE, and 3D-BTE (with bulk phonons combined with the Fuchs-Sondheimer boundary scattering model) methods are presented. The Fuchs-Sondheimer boundary scattering model is applied with specularity parameters $p$ of 0 and 0.34. The green and blue horizontal solid lines indicate the bulk thermal conductivities from 3D-BTE and MD-GK.}
\label{F-thermalconductivity_argon_10K}
\end{figure}

\newpage
\begin{figure}[h]
\centering
\includegraphics{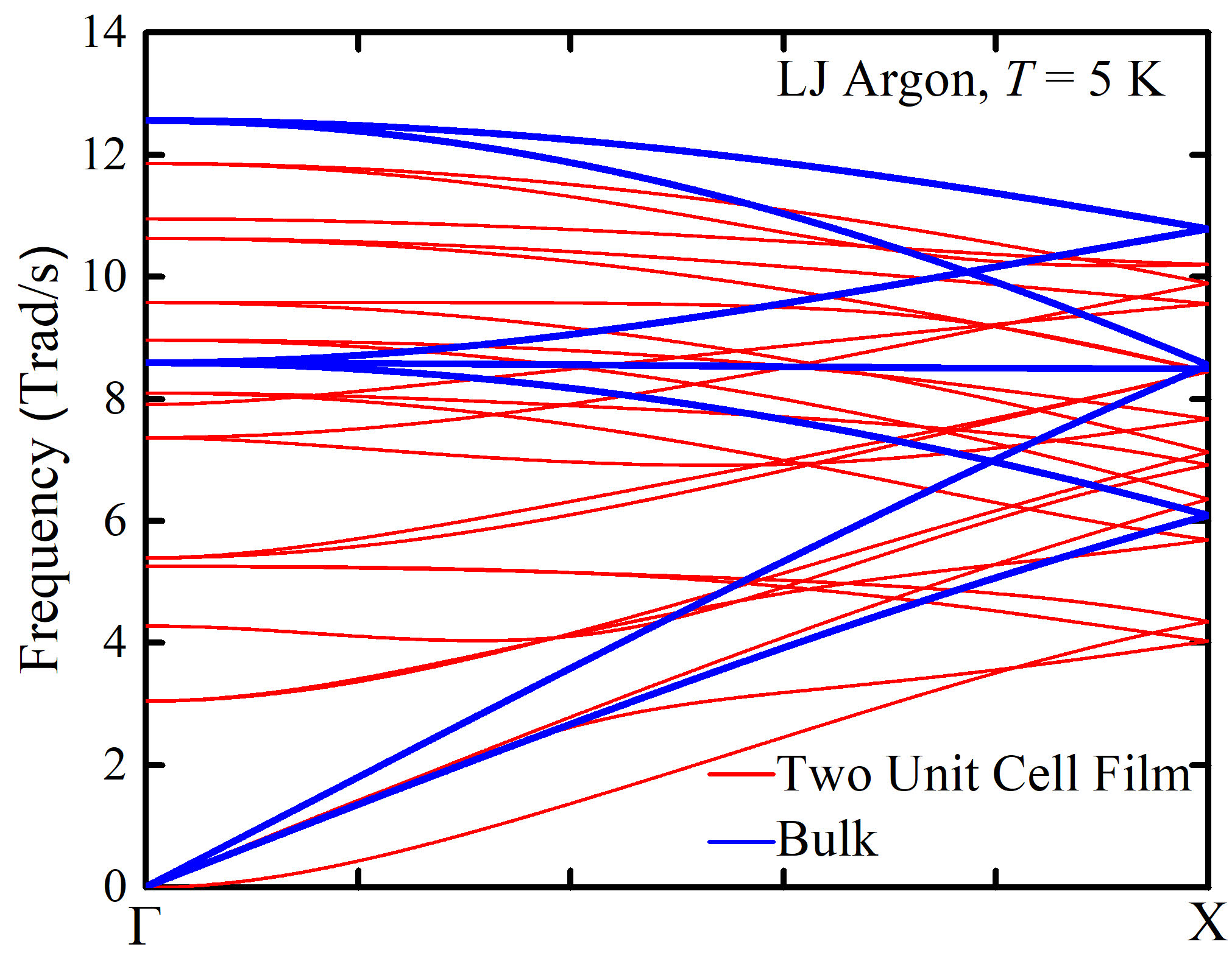}
\caption{[100] phonon dispersion of argon for bulk and the two unit cell film at a temperature of 5 K. The dispersion curves are calculated from harmonic lattice dynamics calculations based on the time-averaged MD structures obtained as described in Sec. III A. The bulk dispersion is based on the four-atom conventional unit cell. The maximum bulk frequency is 12.5 Trad/s, while that for the film is 11.8 Trad/s. 
A flattened phonon branch can be observed at the $\Gamma$-point around 4.9 Trad/s in the two unit cell film, which corresponds to the peak in the density of states in Fig. 6.}
\label{F-argon_disp}
\end{figure}

\newpage
\begin{figure}
    \centering
    \includegraphics{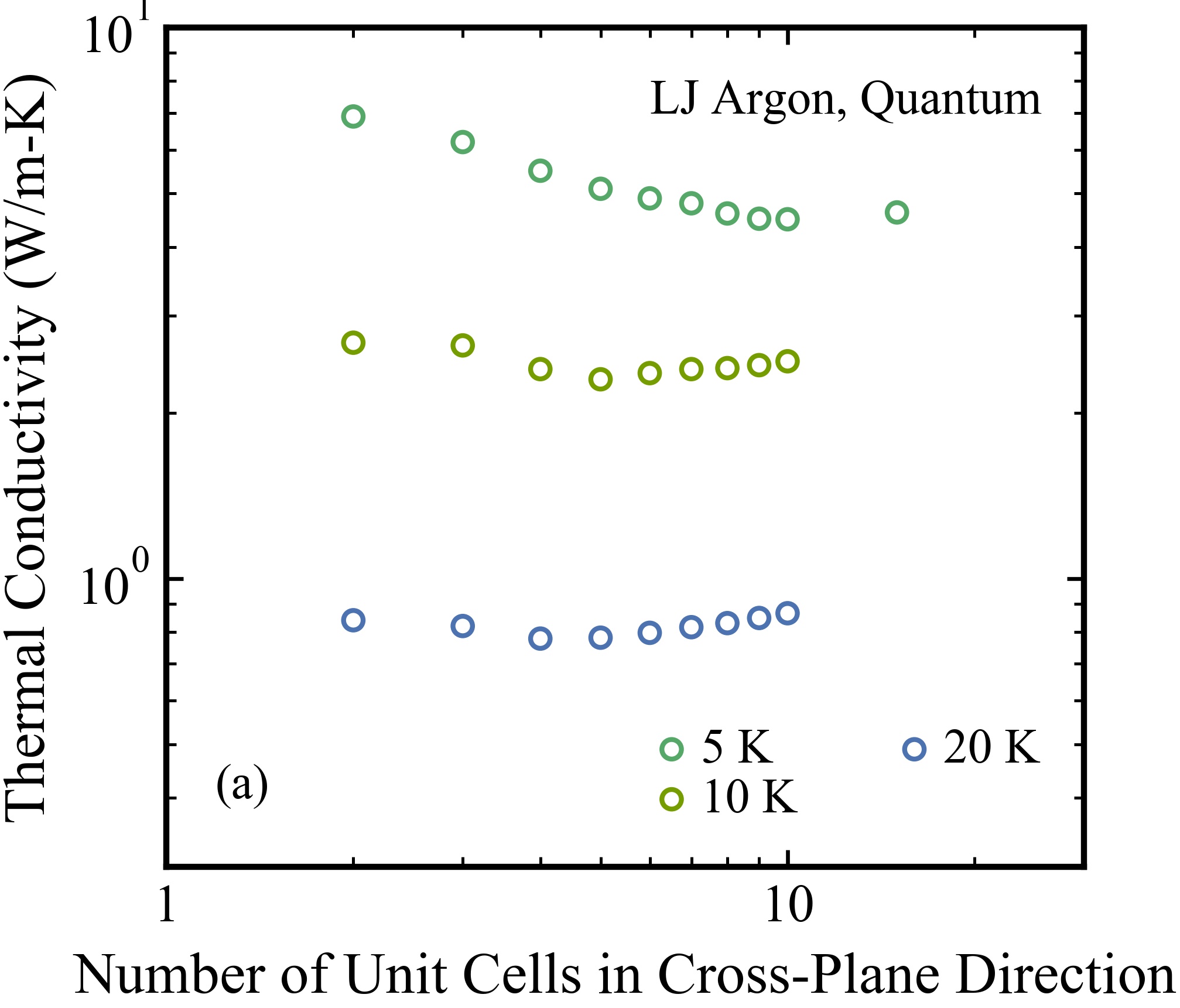}
    \vspace{.3in}\\
    \includegraphics{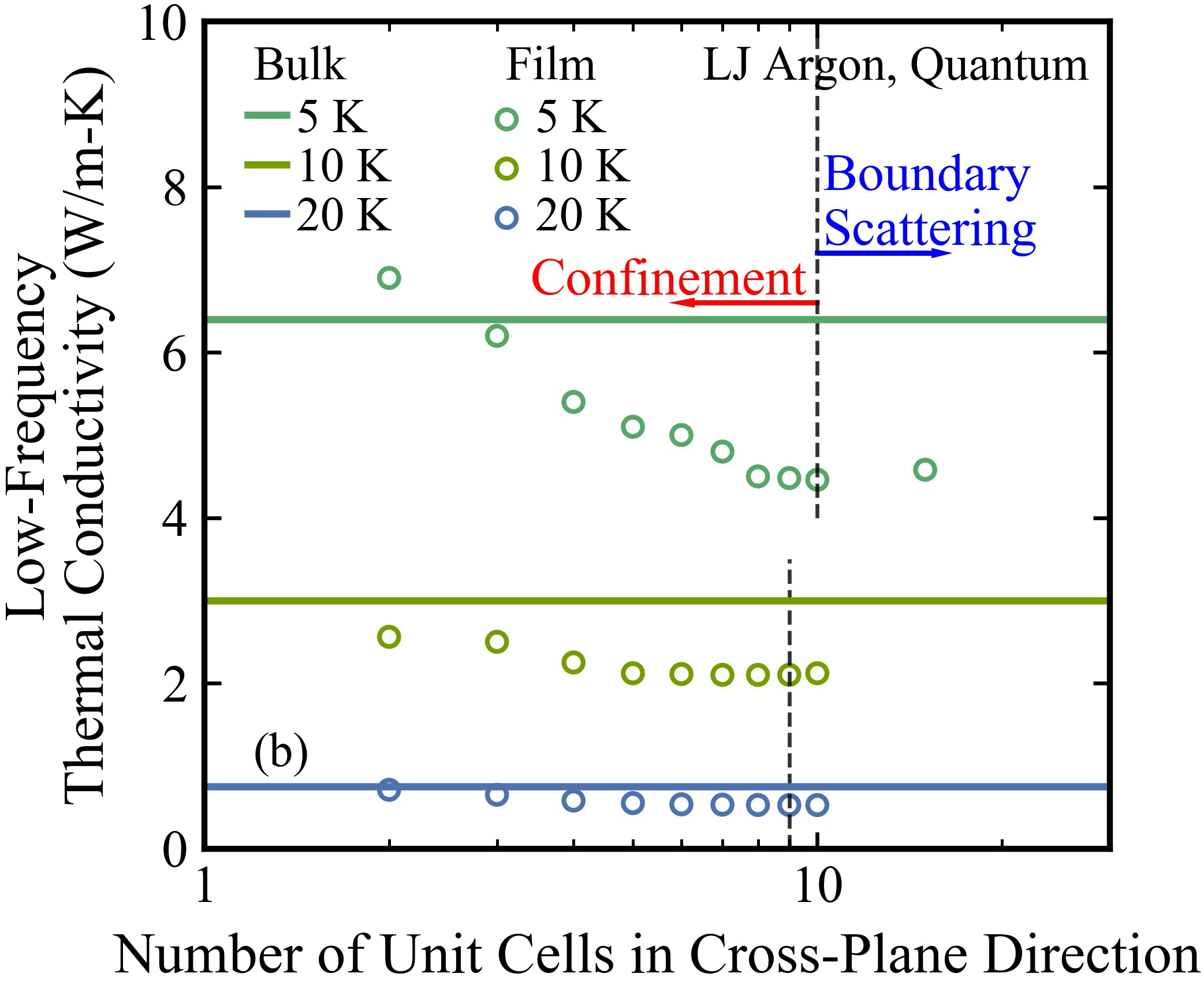}
    \caption{ (a) Thermal conductivity versus thickness for argon films at temperatures of 5, 10 and 20 K. The results are obtained from the 2D-BTE calculations using quantum (i.e., Bose-Einstein) statistics. At 5 K, the total thermal conductivity monotonically increases as the film thickness is decreased below ten unit cells. An increase of the total thermal conductivity is also observed when the thickness is smaller than five unit cells at 10 K and four unit cells at 20 K. These results indicate a stronger confinement effect under quantum statistics compared to classical statistics, where a thermal conductivity plateau was observed in Figs. 5(a) and 5(b). (b) Accumulated thermal conductivity for phonon modes with frequencies below 6 Trad/s plotted versus film thickness. The horizontal solid lines represent the bulk values. The vertical dashed lines indicate the point where the accumulated thermal conductivity starts to increase with decreasing film thickness. This signature of confinement, observed here using quantum statistics, is the same as that found using classical statistics, as shown in Fig. 7(b).
     }
    \label{F-argon_quantum}
\end{figure}

\clearpage
\section{Silicon Results}

\begin{figure}[h]
\centering
\includegraphics{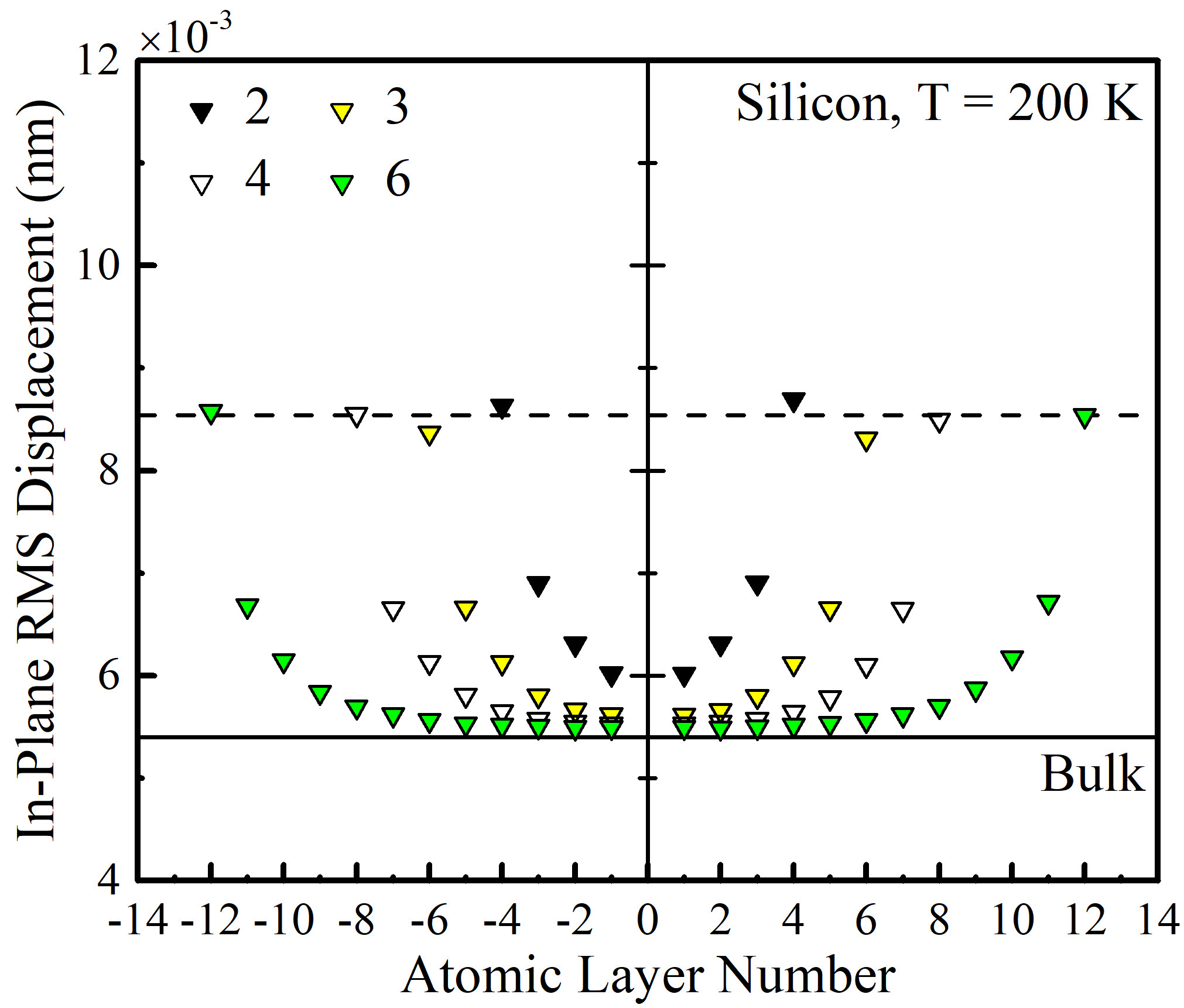}
\caption{In-plane RMS displacement versus atomic layer number for silicon films at a temperature of 200 K. The solid line represents the bulk RMS displacement along one Cartesian direction. The dashed line represents the value that the surface atoms approach in the in-plane direction. The numbers in the legend correspond to the film thickness in conventional unit cells.}
\label{F-rms_in}
\end{figure}

\begin{figure}[h]
\centering
\includegraphics{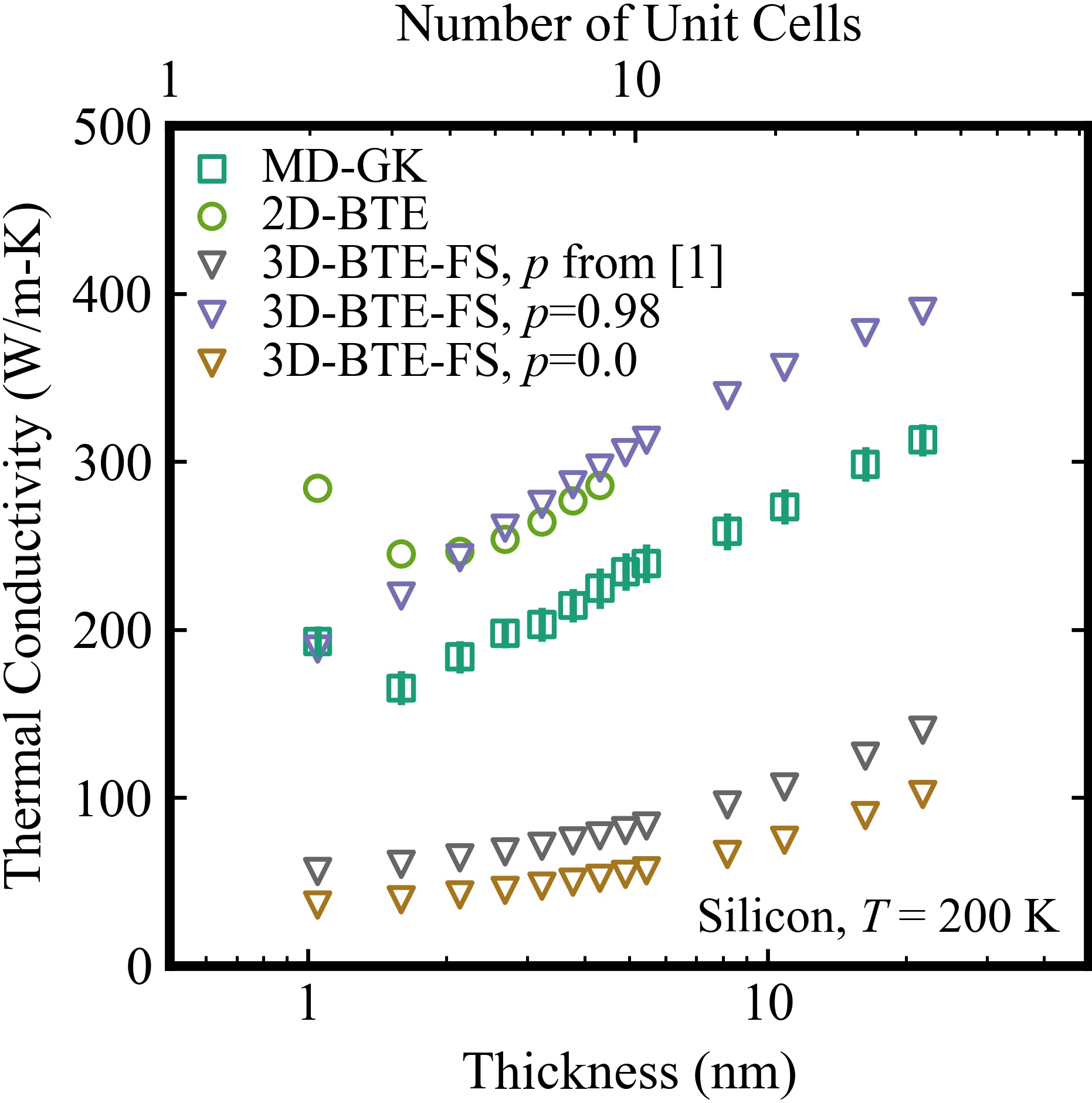}
\caption{Absolute value of thermal conductivity plotted versus thickness for silicon films at a temperature of 200 K. The results from the MD-GK, 2D-BTE, and 3D-BTE (with bulk phonons combined with the Fuchs-Sondheimer boundary scattering model) methods are presented. The Fuchs-Sondheimer boundary scattering model is applied with constant ($p =$ 0.0 and 0.98) or wavelength-dependent specularity parameters \cite{Ravichandran:2018eff}. The bulk thermal conductivity from MD-GK is 414 W/m-K and that from 3D-BTE is 464 W/m-K.}
\label{F-tcabs_silicon}
\end{figure}

\clearpage
\section{\label{sec:crossover}Crossover between 2D-BTE and MD-GK predictions for argon}

As observed in Figs. 5(b) and \ref{F-thermalconductivity_argon_10K}, the MD-GK thermal conductivities unexpectedly become greater than the 2D-BTE values for films thicker than fifteen unit cells at a temperature of 10 K and ten unit cells at a temperature of 20 K. It appears that a crossover will also occur at a temperature of 5 K, but we are unable to observe it due to computational limitations. Noting that, as expected, the 2D-BTE bulk thermal conductivities are higher than the MD-GK values at all temperatures, a second cross-over must occur at a larger film thickness. As temperature increases, the crossover point shifts to a smaller thickness. 

To probe the origin of the crossover, we explore the potential energy surface experienced by an atom, which can provide insight into the anharmonicity that it experiences \cite{parrish2014origins}. 
To do so, a single atom in the time-averaged structure obtained from the MD simulations is displaced by an amount $r$. For bulk, the displacement is chosen to be along the [100] direction. For the films, the displacement is chosen to be for a surface atom in the bottom layer along the cross-plane direction, with a positive value corresponding to motion towards the film center.

\begin{figure}[t]
\centering
\begin{minipage}[t]{0.48\textwidth}
\centering
\includegraphics{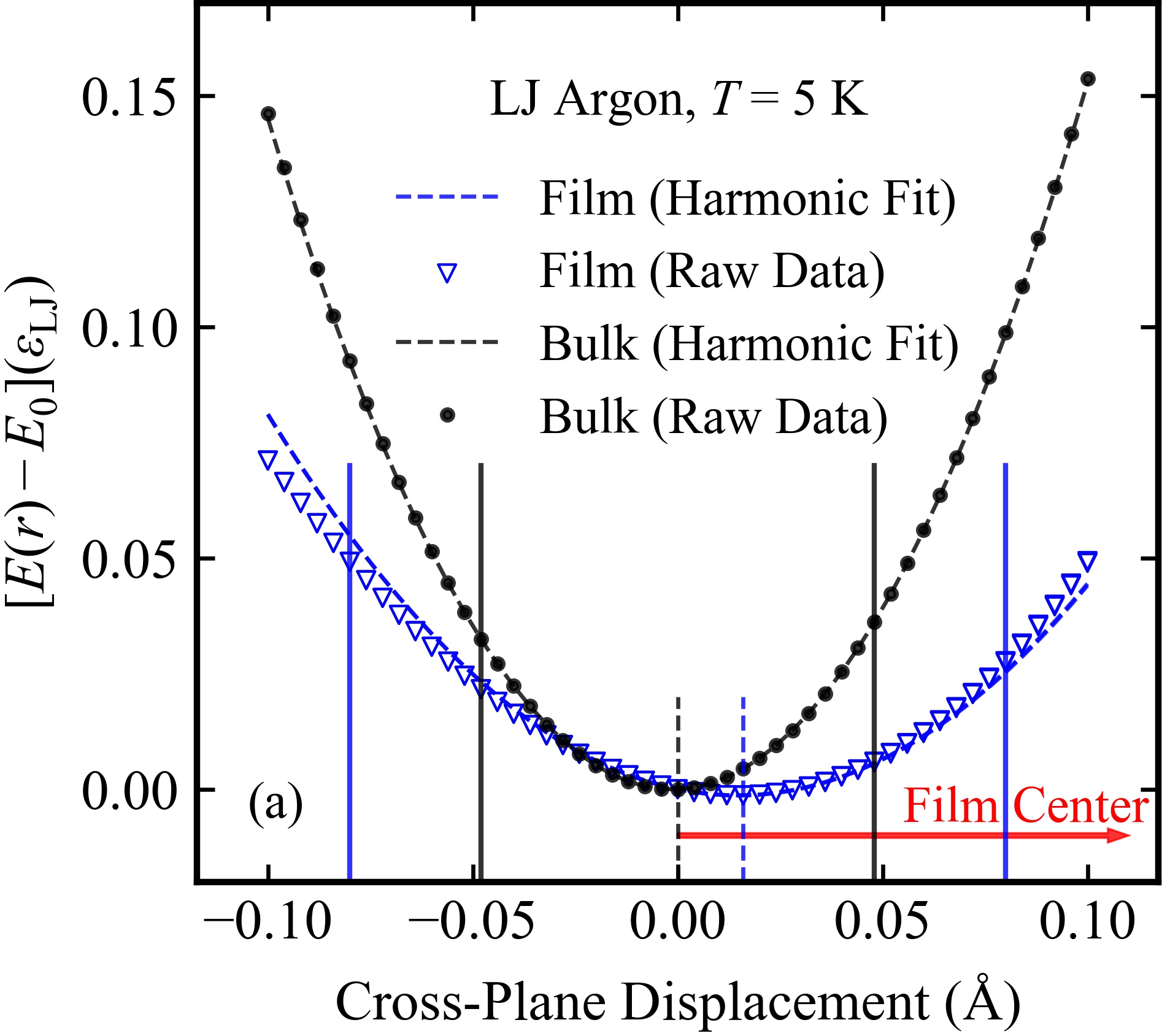}
\end{minipage}
\hspace{.1in}
\begin{minipage}[t]{0.48\textwidth}
\centering
\includegraphics{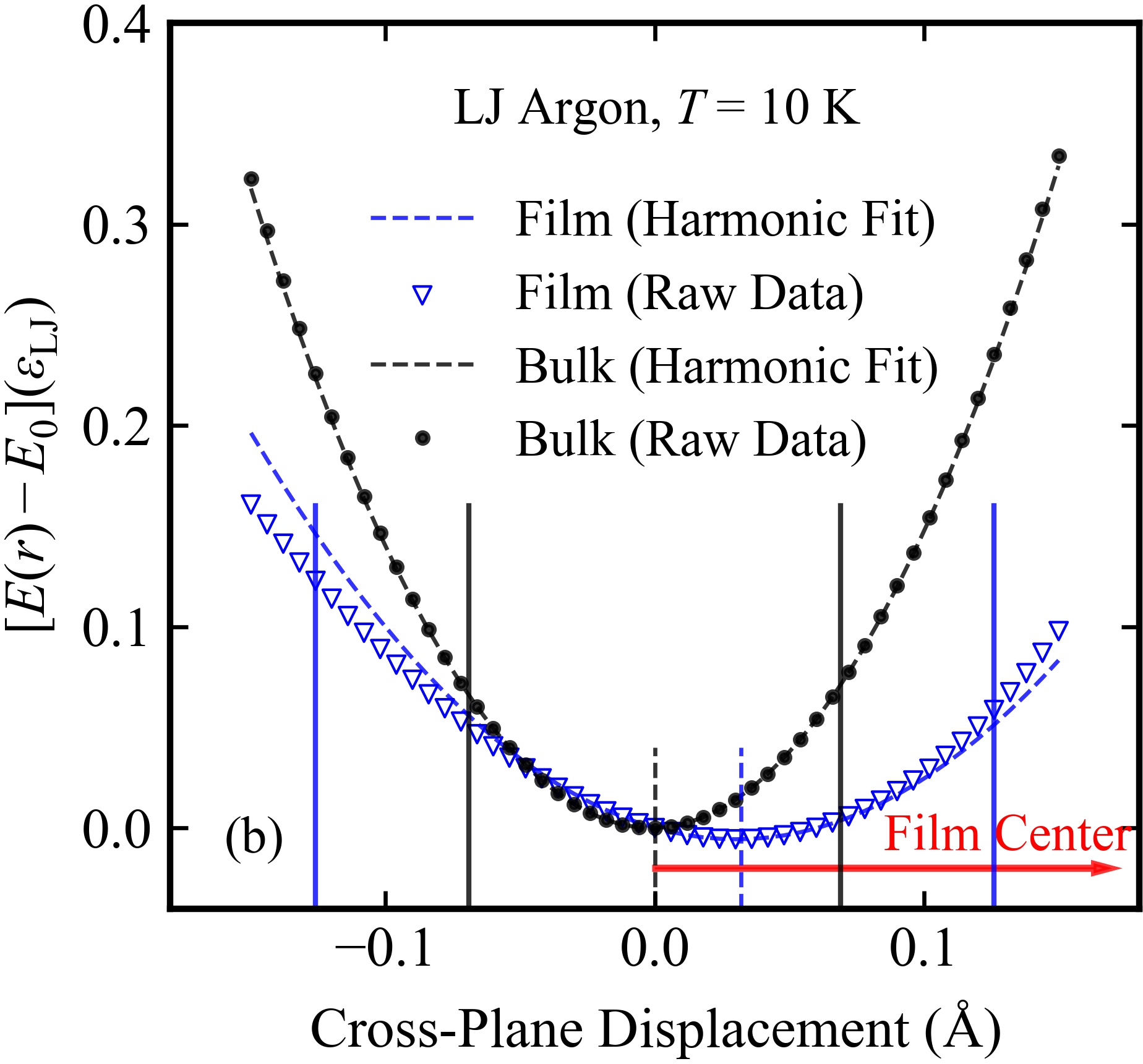}
\end{minipage}\\
\vspace{.1in}
\begin{minipage}[t]{0.48\textwidth}
\centering
\includegraphics{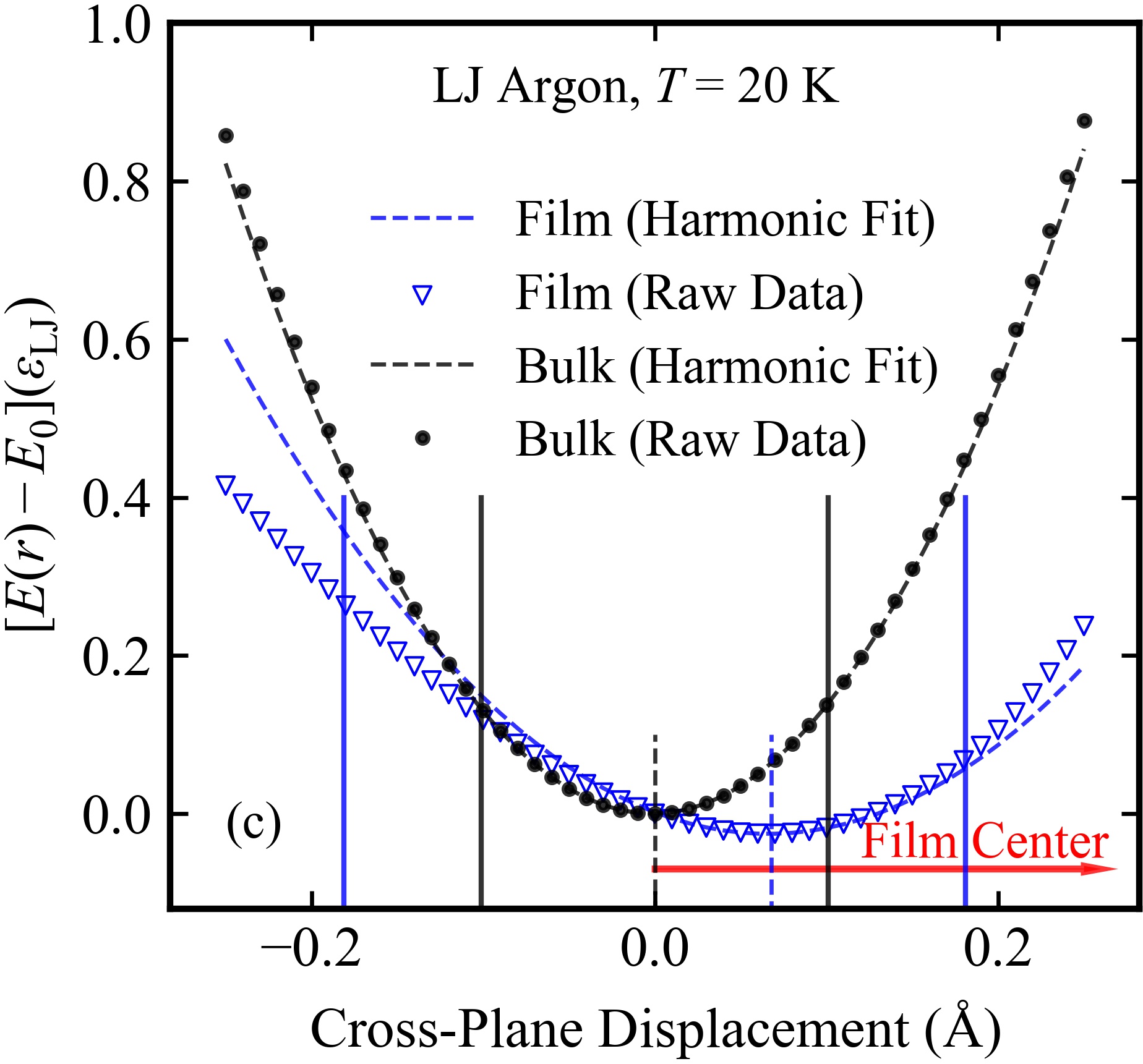}
\end{minipage}
\hspace{.1in}
\begin{minipage}[t]{0.48\textwidth}
\centering
\includegraphics{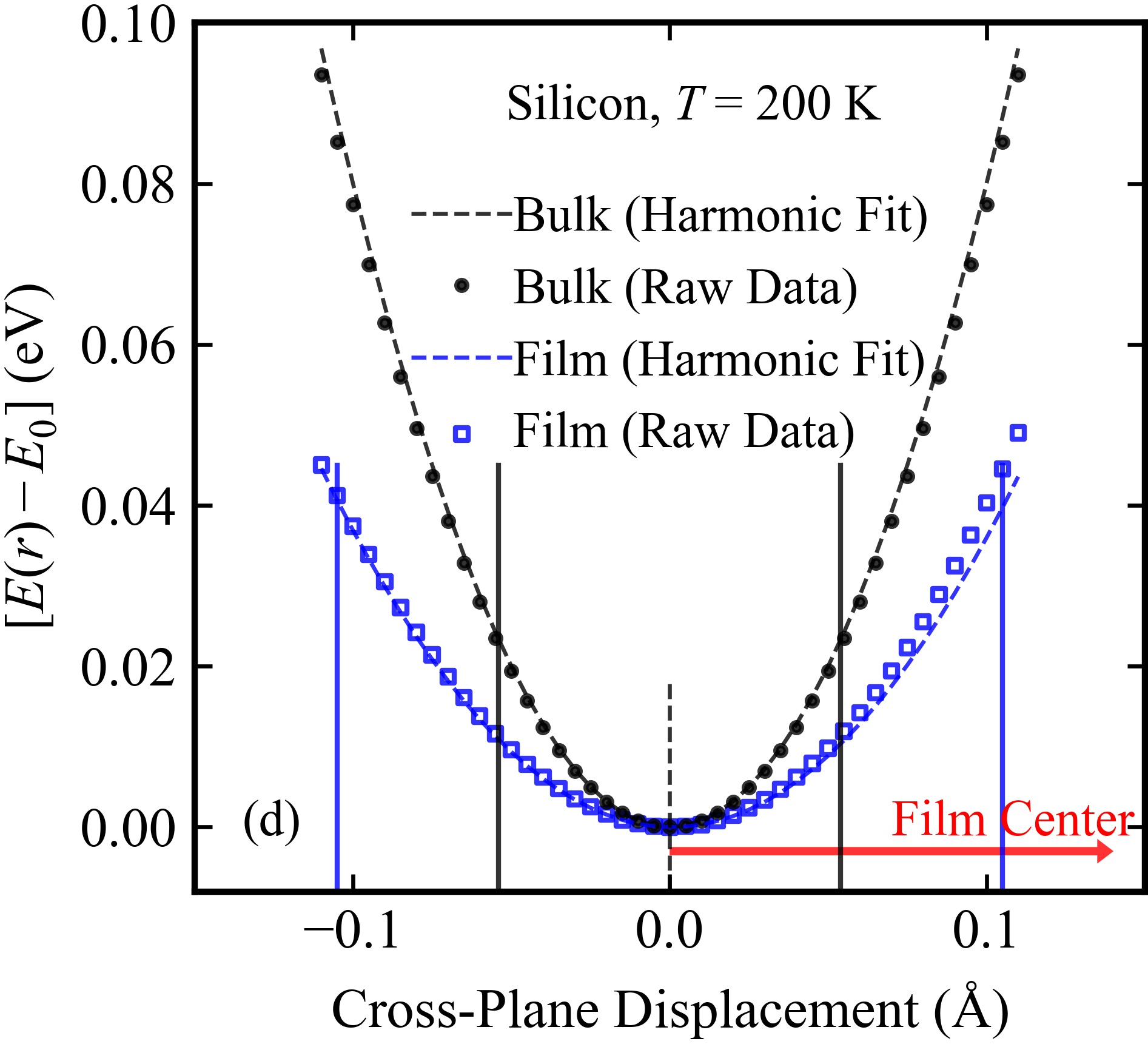}
\end{minipage}
\caption{Potential energy change due to the displacement of one surface atom in the cross-plane direction for the two unit cell argon film at temperatures of (a) 5 K, (b) 10 K and (c) 20 K, and (d) for the two unit cell silicon film at a temperature of 200 K. A positive displacement denotes moving the atom in the bottom layer towards the film center. Also plotted are results for moving an atom in the bulk phase in the [100] direction.
A parabola, plotted as a dashed line, is fit to the first four points of each data set around its energy minimum, which is denoted by a vertical dash line. The horizontal solid blue lines represent the RMS displacement of the surface atom in the cross-plane direction. The horizontal solid black lines represent the RMS displacement of a bulk atom along one of the Cartesian directions.}
\label{F-potential_well}
\end{figure}

The resulting change in potential energy as a function of the displacement size, $E(r)-E_0$, is plotted in Figs.~\ref{F-potential_well}(a)-\ref{F-potential_well}(c) for bulk argon and for the argon film with a thickness of two unit cells at temperatures of 5, 10, and 20 K. The film results are similar for other thicknesses. The solid black(blue) vertical lines denote the RMS displacement of the bulk(film surface) atom along the direction of the displacement. At all temperatures, as expected, the bulk data have an energy minimum at the zero-displacement position, which is denoted with a dashed black vertical line, and are symmetrical about that point. A parabola fit to the bulk data at each temperature provides a good fit. The anharmonic contributions to the potential well (i.e., the deviation from the parabolic fit) at the bulk RMS displacement is less than 2\% for all temperatures.

For the films, however, the change in potential energy has a minimum, denoted by a dashed blue vertical line, that is shifted to the right of the zero-displacement position (i.e., towards the film center). 
The shift increases from 0.016 to 0.032 to 0.066 $\angstrom$ as the temperature increases from 5 to 10 to 20 K. Furthermore, the film data at each temperature are strongly asymmetrical about their minimum, pointing to stronger anharmonicity compared to bulk at the same temperature. This conclusion was also reached through analysis of Figs.~4, \ref{F-rms_all_argon_1020K}(a), and \ref{F-rms_all_argon_1020K}(b).
When moving the surface atom by its RMS displacement, the anharmonic contribution to the potential energy at a temperature of 5 K is 8\% for motion towards the film center and -12\% for motion away from the film center. The anharmonicity increases with increasing temperature, reaching 17\% (towards film center) and -34\% (away from film center) at a temperature of 20 K. 

The perfect (i.e., zero displacement) structures are obtained by time-averaging from MD simulations, as discussed in Sec. II B. 
The shift of the energy minimum and the asymmetrical well indicate that the time-averaged atomic positions do not correspond to the energy minimum. The magnitude of the shift scaled by the RMS displacement increases from 0.20 to 0.25 to 0.36 as the temperature is increased from 5 to 10 to 20 K. 
While the shift of the energy minimum is smaller than the RMS displacement, it is larger than the 0.005 $\angstrom$ displacements used to obtain the harmonic and cubic force constants needed for the 2D-BTE calculations. The force constants that we use are thus an approximation of the true force constants. We believe that this approximation is the origin of the thermal conductivity crossover.
As temperature increases and with it anharmonicity, the crossover point moves to smaller thicknesses.
While it is thus difficult to maintain complete consistency between the 2D-BTE and MD-GK calculations for argon, we believe that it is still appropriate to compare their thermal conductivity predictions.

The potential energy wells experienced by a surface atom in the silicon film with a thickness of two unit cells and by an atom in bulk silicon are plotted in Fig.~\ref{F-potential_well}(d). 
Unlike in the argon films, there is no shift in the location of the energy minimum for the film compared to bulk, which we attribute to the stiffer structure. The force constants used in the 2D-BTE calculations will thus be consistent with the MD-GK simulations.

The energy well of the bulk silicon atom is within 3\% of the harmonic fit at the bulk RMS displacement, pointing to weak anharmonicity. In moving from from bulk to the film, the well curvature decreases and an asymmetry emerges. The anharmonic contributions are always positive. At the RMS displacement, they are 12\% and 1\% when moving towards and away from the film center, which is a smaller effect than seen for argon at all temperatures considered.
Therefore, as the film thickness decreases, the increased deviation of thermal conductivity between MD-GK and 2D-BTE shown in Fig. 10 is consistent with the interpretation of the increased anharmonicity discussed in Sec. IV B. Furthermore, no crossover of thermal conductivity between MD-GK and 2D-BTE is observed in the silicon films.

\bibliographystyle{apsrev4-1} 
\bibliography{Ref0303.bib}